\documentclass[12pt,preprint]{aastex}

\usepackage{wasysym}
\usepackage{amssymb}

\def\arcsecf {\hbox{$.\!\!^{\prime\prime}$}}

\shorttitle{Turbulence Profiles}
\shortauthors{Els et al.}

\begin{document}

\title{Thirty Meter Telescope Site Testing VI: \\ Turbulence Profiles}


\author{S.G. Els\altaffilmark{1,2}, T. Travouillon\altaffilmark{2}, M. Sch\"ock\altaffilmark{3}, 
R. Riddle\altaffilmark{2}, W. Skidmore\altaffilmark{2}, \\ J. Seguel\altaffilmark{1}, E. Bustos\altaffilmark{1}, D. Walker\altaffilmark{1}  }

\altaffiltext{1}{Cerro Tololo Inter-American Observatory, National Optical Astronomy Observatory, Casilla 603, La Serena, Chile}
\altaffiltext{2}{TMT Observatory Corporation, 2632 E. Washington Blvd., Pasadena CA 91107, USA}
\altaffiltext{3}{TMT Observatory Corporation, 5071 West Saanich Road, Victoria, British Columbia, Canada }

\begin{abstract}
The results on the vertical distribution of optical turbulence above the five mountains which were 
investigated by the site testing for the Thirty Meter Telescope (TMT) are reported. On 
San Pedro M\'artir in Mexico, the 13 North site on Mauna Kea and three mountains in northern Chile 
Cerro Tolar, Cerro Armazones and Cerro Tolonchar, MASS-DIMM turbulence profilers have been operated over 
at least two years. Acoustic turbulence profilers -- SODARs -- were also operated at these sites. 
The obtained turbulence profiles indicate that at all sites the lowest 200~m are the main source of the 
total seeing observed, with the Chilean sites showing a weaker ground layer than the other 
two sites. The two northern hemisphere sites have weaker turbulence at altitudes above 500~m, with 13N 
showing the weakest 16~km turbulence, being responsible for the large isoplanatic angle at this site. 
The influence of the jetstream and wind speeds close to the ground on the clear sky turbulence strength 
throughout the atmosphere are discussed, as well as seasonal and nocturnal variations. 
This is the sixth article in a series discussing the TMT site testing project.

\end{abstract}
\section{Introduction}
The TMT project conducted an extensive site monitoring campaign to identify the most suitable site to host the observatory. 
Knowing the vertical distribution of the optical turbulence strength (TS), as expressed by the 
structure constant of the refractive index $C_n^2$, in Earth's atmosphere -- the turbulence profile (TP or $C_n^2(h)$) -- 
is of great importance for modern astronomical observing techniques. Especially adaptive optics (AO) instrumentation in its various forms relies in 
its inital design on a good assumption of the typical altitude distribution of the TS above the telescope. AO will become an important observing 
technique of the future extremely large telescopes (ELTs) and it is therefore essential to obtain measurements of the TPs above the ELT 
candidate sites. Such measurements can in principle be obtained in-situ by means of balloon borne turbulence sensors (\citealt{barletti77}). 
But a continuous monitoring of the TP using balloons would be very costly. Therefore, site testing programs use remote sensing techniques 
to obtain measures of the total TS. The instruments used during such campaigns have to be rugged and should only require small size telescopes. 
These requirements limited early site testing programes to measurements of the total seeing only and were first based on visual observations 
(\citealt{stock64}), photoelectric (\citealt{irwin66}) or photographic techniques (\citealt{birkle76}). A radical improvement of measuring 
the total seeing was obtained by the development of the Differential Image Motion Monitor (DIMM, \citealt{sarazin}). Remote sensing 
of the actual TP based on measurements of stellar scintillations has been done since some time by means of the SCIntillation Detection And Ranging (SCIDAR) 
technique (\citealt{vernin73}). This technique, however, requires telescopes with apertures of at least 1~m for being effective. Therefore, in recent years 
several other techniques, like SLOpe Detection And Ranging (SLODAR, \citealt{butterley06}) or Multi Aperture Scintillation Sensor (MASS, \citealt{toko02}) 
have been developed to overcome the requirement for such large apertures and still obtain TPs with reasonable altitude coverage and resolution. The TMT site 
testing employed combined MASS-DIMM instruments (\citealt{kornilov07}) to assess the low resolution TP through the entire atmosphere. 
In order to measure the TP within the lowest few hundred meters above the sites, acoustic turbulence profilers SODARs (SOund Detection And Ranging, 
\citealt{crescenti97}) were also employed.  

Here we report on the results obtained with these turbulence profilers during the TMT site testing. 
In Section~\ref{sitesdata} the candidate sites, the deployed instrumentation and data are described. Sections~\ref{massbasedprofiles} 
and \ref{sodarbasedprofiles} show and discuss the observed ``typical'' TPs. Sections~\ref{seasonalprofiles}, \ref{nightprofiles} and \ref{meteorelations} 
discuss seasonal and nighttime changes and the dependence on the ground wind speed of the TS at the different altitudes. 

\section{TMT site testing data}
\label{sitesdata}
After a preselection based on satellite data and general topographic conditions (\citealt{walker09}) , TMT deployed 
site testing stations on five candidate sites: 13N on Mauna Kea (Hawaii), San Pedro M\'artir or SPM (Mexico), 
Cerro Tolar, Cerro Armazones and Cerro Tolonchar (all in northern Chile). The general properties 
of these mountains are summarized in Table~\ref{sites}. More detailed information about these sites is given in \cite{schoeck07} and \cite{schoeck08}. 
These stations have been operated for at least two years on each mountain, with a considerable time span 
being simultaneous between all stations (see Table~\ref{sites}). 

All TMT site testing stations were equipped with a custom built 35~cm Cassegrain telescope, manufactured by 
Teleskoptechnik Halfmann. These telescopes were mounted on 6.5~m tall towers, resulting in an elevation of the 
telescopes of approximately 7~m above ground. All components of the TMT site testing equipment are summarized in \cite{schoeck07}
and \cite{riddle08}. 
On each telescope a combined MASS-DIMM device (\citealt{kornilov07}) was deployed. The DIMM channel of these instruments measures the differential 
image motion of the star, which then results in the 
total seeing from the telescope level to the top of the atmosphere (\citealt{sarazin}). 
The used DIMM system and its precision is described in detail by \cite{wang07}, who reported the seeing 
precision -- the comparability between the TMT DIMMs -- to be better than 0\arcsecf02. 

The MASS (Multi Aperture Scintillation Sensor) reconstructs, by measuring the spatial structure of the flying shadows, 
a TP at six altitudes $h_{i=1...6}=0.5, 1, 2, 4, 8, 16~$km above the telescope (\citealt{toko07}). 
The MASS TPs consist of the integrals of $C_n^2$ over each altitude bin, weighted with 
functions which peak at the the layer height $h_i$ and go to zero at the neighboring MASS layer altitudes (thus causing 
some overlap between neighboring layers). 
Therefore, the MASS TPs are given in $C_n^2(h)dh$, which is equivalent to the seeing of the layer to the power of $(5/3)$.
From such a profile the seeing which would be seen by an observer 500~m above the telescope to the top of the atmosphere 
is computed. Also the isoplanatic angle $\theta_0 \propto \sum_i(C_n^2(h)h^{5/3}dh)^{-3/5}$ can directly be obtained from 
these profiles. The precision of the strength of the individual MASS layers was found by \cite{els08} to be better than 
$10^{-14}$~m$^{1/3}$ and of the MASS seeing to be better than 0\arcsecf05. 

All turbulence values in this paper are given at a wavelength of $\lambda=500~$nm and corrected for zenith distance. 

From the difference between MASS and DIMM seeing data, the seeing originating between the site testing telescope and 
the first MASS layer can be computed by 
\begin{equation}
\epsilon_{\mathrm{GL}} = (\epsilon_{\mathrm{DIMM}}^{5/3} - \epsilon_{\mathrm{MASS}}^{5/3})^{3/5} .
\label{glformula}
\end{equation}
We refer to this lowest layer seeing as ground layer seeing (GL). In order to obtain the GL, only MASS data 
were used for which DIMM data exist within 60~seconds. This is usually the case as the DIMM channel 
is used to acquire the target star (see below). When the GL is so weak that its strength is comparable to the 
measurement accuracy of MASS and DIMM, the MASS seeing can become larger than the DIMM seeing. We treat these cases 
by applying the exponent ($3/5$) in eq.~\ref{glformula} on the absolute of the difference and reverting the sign of 
the result in case $\epsilon_{\mathrm{DIMM}} < \epsilon_{\mathrm{MASS}}$, same as in \cite{skidmore08}.
Even though negative GL values are unphysical, not taking them into account would bias the resulting statistics; 
in particular the low seeing percentiles. 

The atmospheric turbulence time constant $\tau_0\propto \sum_i(C_n^2(h)v(h)^{5/3}dh)^{-3/5}$ requires the 
additional knowledge of the vertical wind speed profile $v(h)$ at the time of observation, which is usually unknown. 
However, \cite{toko02} found that MASS data can in principle be used to infer a $\tau_{0,\mathrm{MASS}}$ value based on 
the differential exposure time scintillation index. But this $\tau_{0,\mathrm{MASS}}$ has to be properly calibrated to 
obtain the real $\tau_0$ and is generally unreliable for $\tau_0 \gtrsim 5~$ms in the case of the TMT MASS systems (\citealt{els08}).  
A thorough  investigation of the $\tau_0$ measurements obtained by the TMT site testing is given 
by \cite{travouillon08c}. Here, we make use of the uncalibrated $\tau_{0,\mathrm{MASS}}$ data but only for 
data selection purposes for specific analyses within Section~\ref{massbasedprofiles}. 

MASS and DIMM measurements are triggered simultaneously and the DIMM integrates the stellar light for 36~sec, whereas the 
MASS samples 60~sec. Including processing time, this results in a MASS and DIMM measurement every 70--90~sec. 
As the DIMM CCD is used to acquire the target star, typically for more than 95\% of the collected MASS data a DIMM measurement 
was taken within 60~sec. However, sometimes the DIMM or MASS computations failed to provide valid results. This is in part due to 
data taken under cloudy conditions. As the robotic system operated the telescope whenever the environmental conditions did allow 
a safe operation, a high data collection efficiency was assured (\citealt{riddle06}).  
Nightly operations commenced approximately one hour after sunset and ceased approximately one hour before sunrise. 

The TMT site testing project also operated acoustic turbulence profilers SODARs (SOund Detection And Ranging, i.e., \citealt{crescenti97}) 
at the candidate sites. These systems emit sound pulses into the atmosphere and obtain from the backscatter signal, the turbulence and wind 
speed profiles above the site. Two SODAR systems, SFAS and XFAS (both manufactured by SCINTEC), were operated. The SFAS profiler, measures from 10~m 
to 200~m above the sites with an altitude resolution of 5~m. XFAS profiler range from 40~m to 800~m with a resolution of 20~m. 
The TMT site testing project operated three SFAS and XFAS pairs which were rotated among most sites (no SODAR was operated on Tolar). 
These systems obtained turbulence and wind speed profiles every 30~min. The details on the operations, calibrations and accuracy are given 
in \cite{travouillon08a}. 

\begin{figure}[h]
\caption{Amount of data collected each month of the year at all candidate sites.}
\label{datapts}
\resizebox{0.9 \textwidth}{!}{\rotatebox{270}{\includegraphics{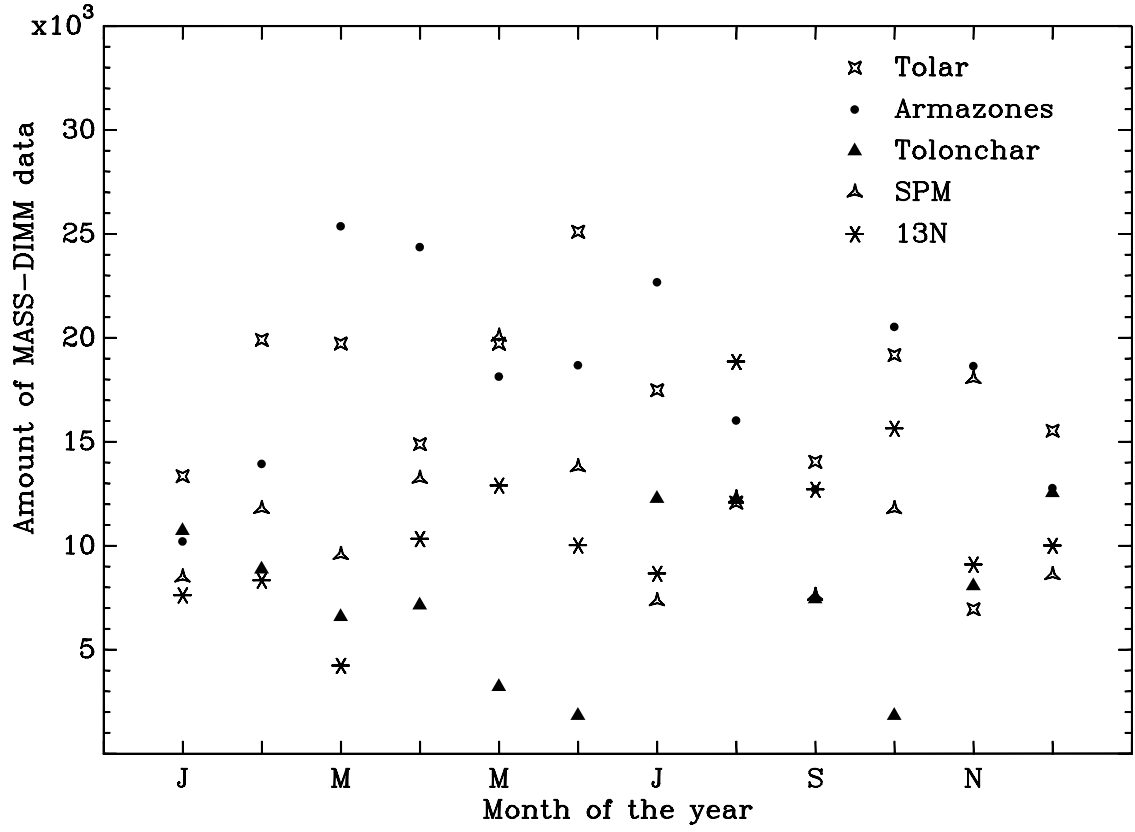}}}
\end{figure}

Table~\ref{sites} indicates the time spans covered by the data shown in this work. Figure~\ref{datapts} shows the amount of 
simultaneous MASS-DIMM data collected at each site during each month of a year. 
Even though data collection was ongoing on several sites until mid 2008, only the data obtained during the indicated periods 
were considered in the TMT site testing process and are presented in this work. The site testing stations at Cerro Tolar, 
Mauna Kea 13N, SPM and Cerro Tolonchar were dismantled in August 2007, June 2008, August 2008 and November 2008, respectively. 
At some of these sites, MASS-DIMM data collection had already been terminated before these dates.

\section{MASS based vertical turbulence profiles}
\label{massbasedprofiles}
The collected MASS observations allow the computation of the characteristic turbulence profile above each of the candidate sites. 
There is, however, no unique definition of a \emph{characteristic} vertical turbulence profile that is ideally suited for all applications. 
In the following we describe two different definitions of representative turbulence profiles. In both methods we include the GL 
layer strength in these profiles and therefore present only data when MASS and DIMM data were available simultaneously. 
As described before, this does not significantly bias or limit the data sample. 

The first way is to compute the median and mean TS for each MASS layer over the entire data set. 
The statistics -- 25\%ile, median and 75\%ile -- of the individual layer TSs are shown in Tab.~\ref{profilestable}. 
For completeness that table also shows the mean and rms values for each layer. 
These data represent the statistical properties of each individual layer, but these profiles can not be considered as realistic or 
representative. In particular because the layers will not be entirely independent from each other. For example, for the lowest MASS layers, 
which are spaced close to each other,  one might expect that shearing forces could cause interaction between these layers. 

\begin{figure}[h]
\caption{Median turbulence profiles observed above the TMT candidate sites. For values see 
Tables~\ref{profilestable}, \ref{profileDIMMseltable}, \ref{profileTHETA0seltable} and \ref{profileTAU0seltable}.
\emph{Panel a:} Overall median profile; \emph{Panel b:} Median of the 10\% of the profiles 
selected around the median DIMM seeing; \emph{Panel c:} Like panel (b) but selected around 
the median MASS $\theta_0$; \emph{Panel d:} Like panel (b) but selected around the median MASS $\tau_0$.}
\label{massprofiles}
\resizebox{0.45 \textwidth}{!}{\rotatebox{0}{\includegraphics{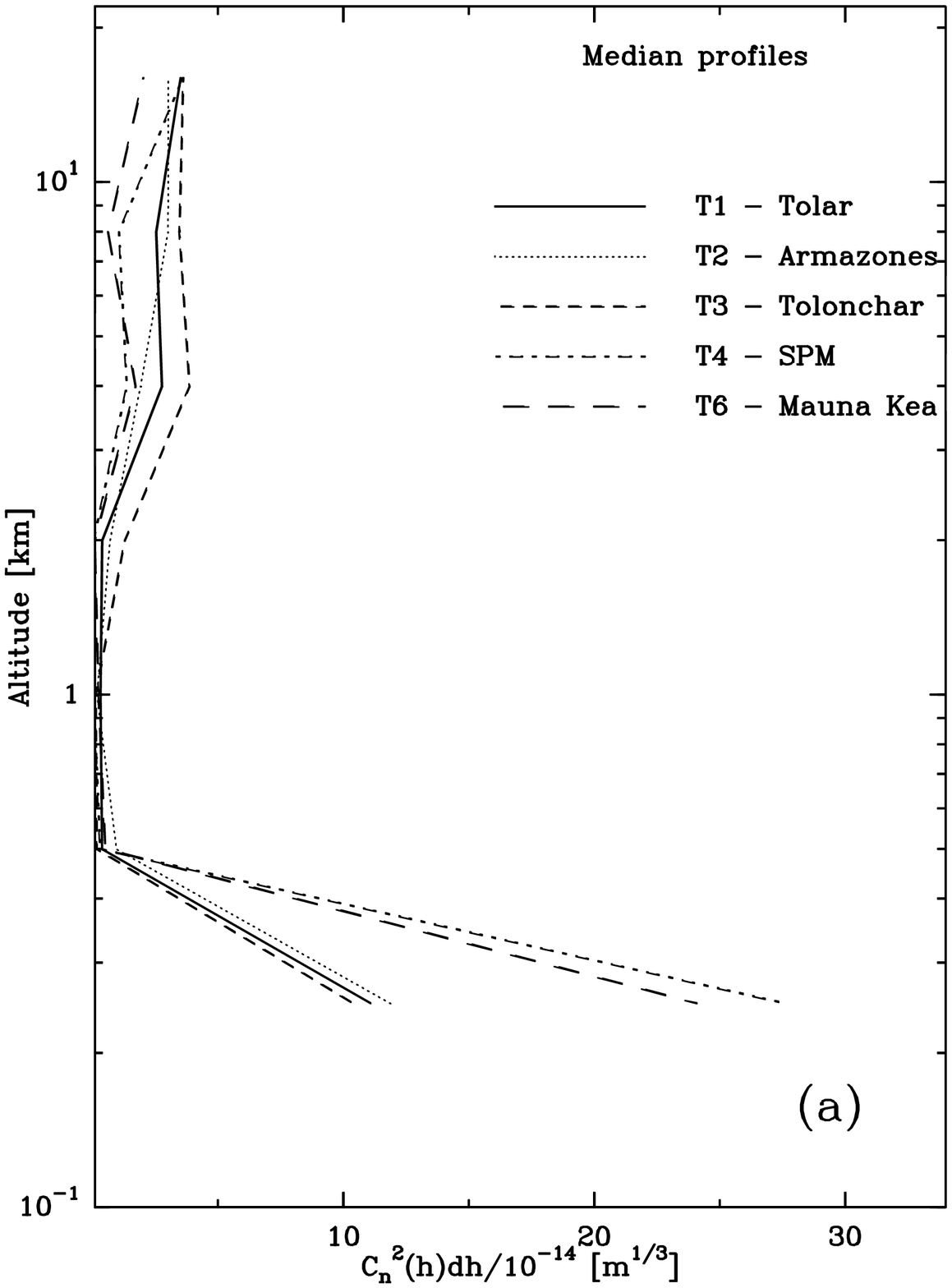}}}
\resizebox{0.45 \textwidth}{!}{\rotatebox{0}{\includegraphics{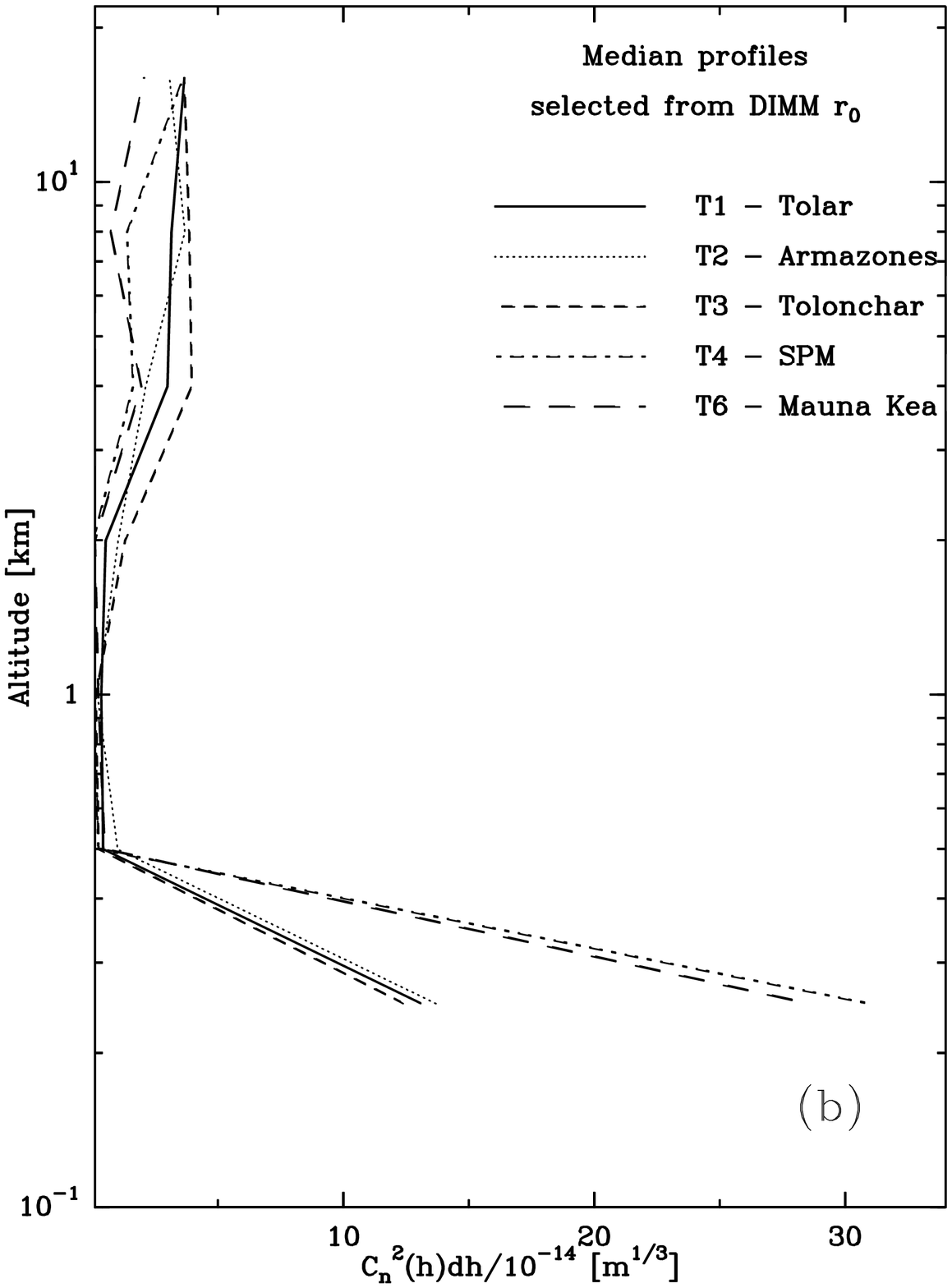}}}
\resizebox{0.45 \textwidth}{!}{\rotatebox{0}{\includegraphics{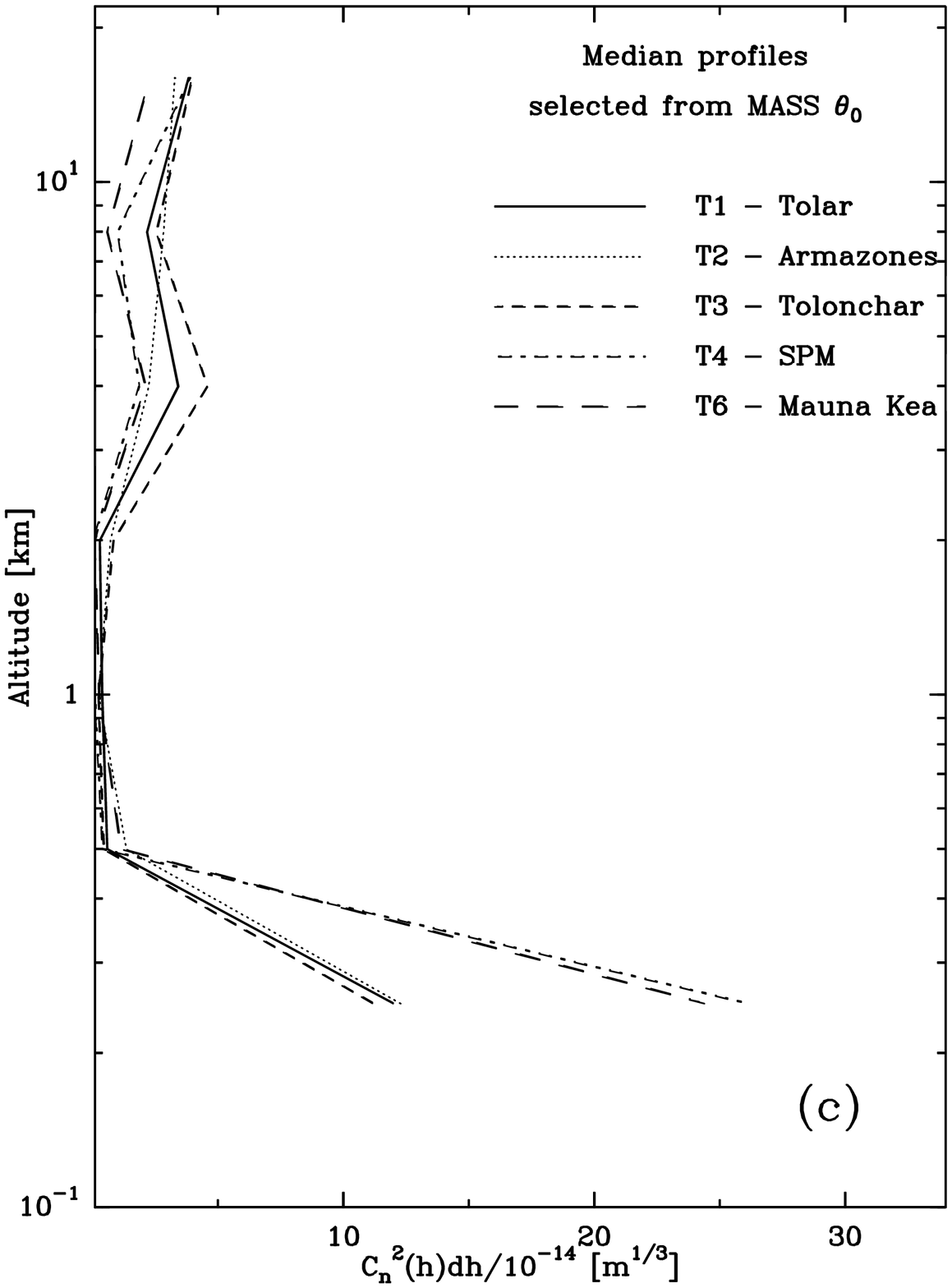}}}
\resizebox{0.45 \textwidth}{!}{\rotatebox{0}{\includegraphics{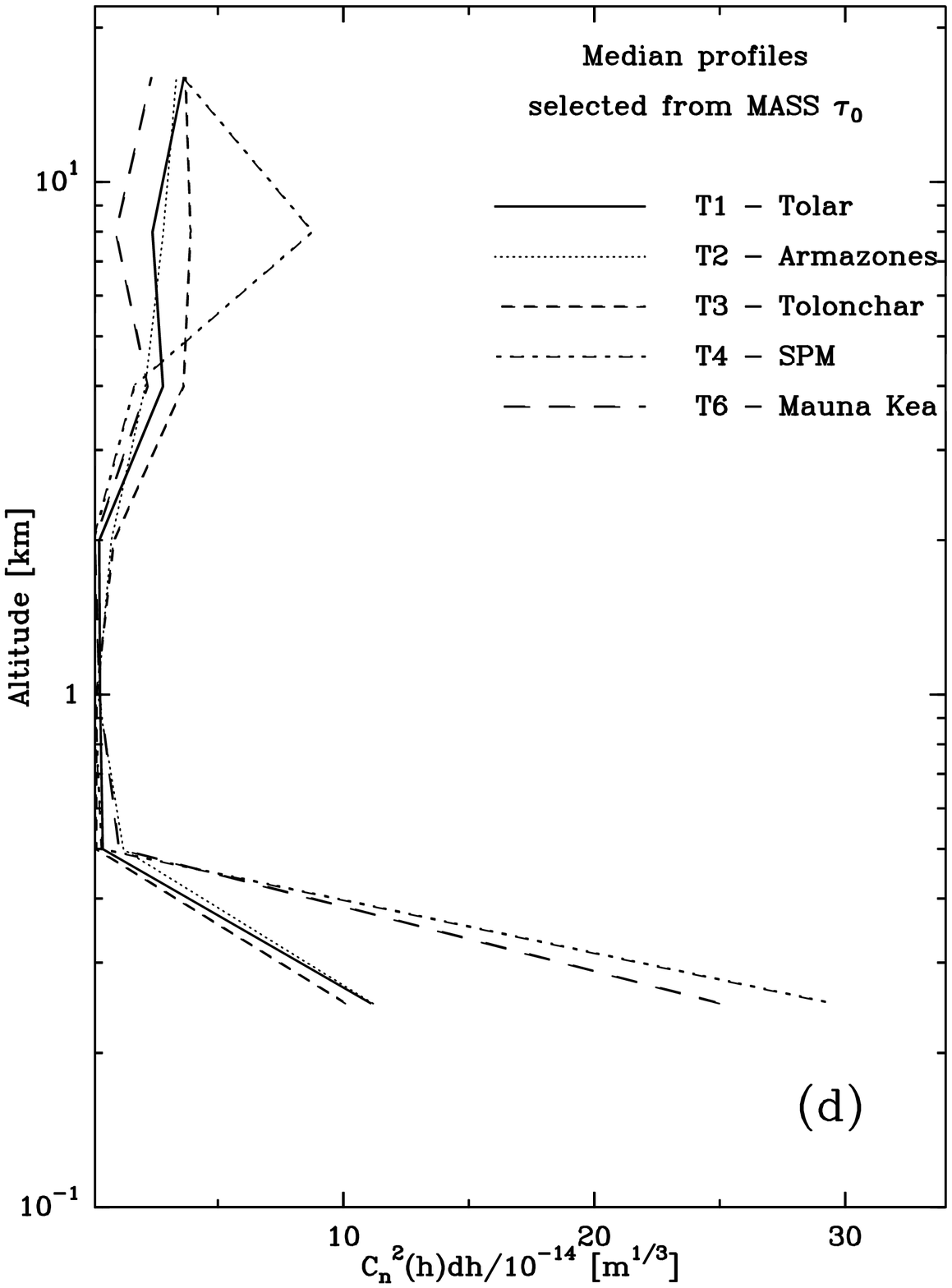}}}
\end{figure}

The other way to define a TP is to compute the TS of the layers under particular seeing, isoplanatic angle and coherence time conditions.  
For example, we looked for the typical TP for the $x$\%ile of the DIMM seeing, which is inverse proportional to the Fried parameter $r_{0,x}$. 
To this end, we first identified $y$\% of all DIMM data with values closest to $r_{0,x}$. Then, we compute the statistics of the MASS data which were 
taken simultaneous with the selected DIMM measurements. In this study we set $y=10$~\%, e.g., $\pm5$~\% of all data around the 25, 50 and 75 DIMM seeing 
percentiles were used to obtain the turbulence profile. The so defined profiles for the median $r_0, \theta_0$ and $\tau_0$ 
conditions are shown in the panels b, c and d of Figure~\ref{massprofiles}. For reference, Tables~\ref{profileDIMMseltable}, \ref{profileTHETA0seltable} 
and \ref{profileTAU0seltable} in the Appendix, also show the mean and  medians of the profiles computed around the 25, 50 and 75 percentiles 
of respective integrated turbulence parameter.

It must be noted that even though these profiles represent the typical vertical distribution of turbulence during the times when a certain value of an 
integrated turbulence parameter is observed,  computing that parameter from these profiles does not result in the value of the 
respective percentile of that parameter computed from the overall data sample. For example in the case of Tolar, computing the 
total seeing from the median profile from MASS data selected around the median DIMM seeing results in 0\arcsecf54. But the overall 
DIMM seeing statistics results in a median seeing of 0\arcsecf63 (\citealt{skidmore08}). This is because in the 
first case the statistics is done before the exponent of $3/5$ is applied to the layer TS, whereas the median seeing is obtained after that step.

These turbulence profiles now allow to understand the differences in the integrated turbulence parameters between the sites (\citealt{skidmore08}). 
The three sites in Chile show a very similar vertical distribution of the TS. 
Compared to the northern hemisphere sites, 13N on Mauna Kea and SPM, the Chilean sites show a weaker ground layer TS 
by approximately 50\%. However, the TS above the Chilean sites is stronger than the northern sites at altitudes above 4~km. 
At these altitudes 13N shows the weakest TS, even less than SPM whose 16~km layer is approximately as strong as that of the Chilean sites. 
This explains why SPM shows a MASS seeing  similarly low as 13N but an isoplanatic angle of the same level as the Chilean sites. 
Because strong weight is put on the highest layers in the isoplanatic angle calculation ($\sim h^{5/3}$), this makes the 16~km MASS layer dominant 
for this specific parameter and results in 13N showing the largest isoplanatic angle. For the seeing calculation all altitudes are weighted equally, 
explaining the low MASS seeing of SPM. 

\begin{figure}[h]
\caption{Turbulence profiles observed above 13N on Mauna Kea. These profiles are the medians of the 10\% of all profiles observed 
closest to the times of the indicated percentiles of seeing $r_0$ and isoplanatic angle $\theta_0$. }
\label{t6profiles}
\resizebox{0.9 \textwidth}{!}{\rotatebox{90}{\includegraphics{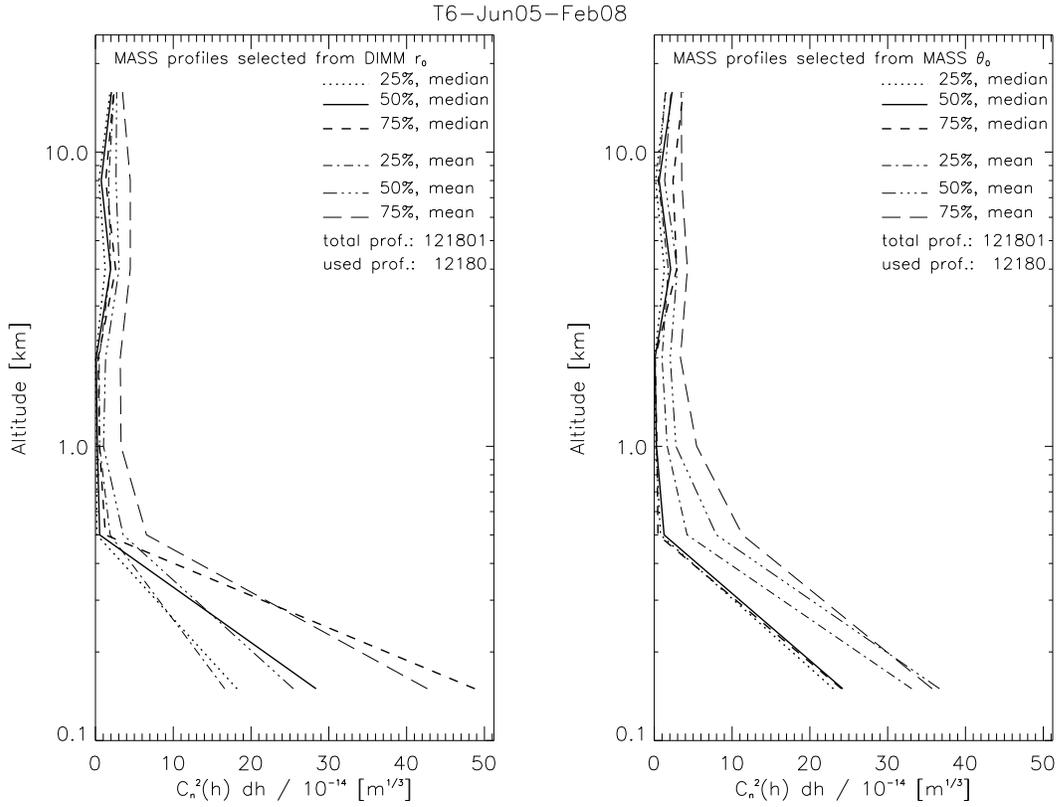}}}
\end{figure}

The different altitude weighting is also visible in the selected turbulence profiles. The spread between the 
profiles selected from the various $r_0$ percentiles is larger than the spread between the $\theta_0$ selected profiles. 
As example, Figure~\ref{t6profiles} shows the profiles from 13N (from Tables~\ref{profileDIMMseltable} and \ref{profileTHETA0seltable}).
It can be seen that the difference between the 25\%ile and 75\% GL TS median values selected from $r_0$ is approximately a factor 
of 30 larger than in the case of the $\theta_0$ selected profiles. Again, the strong weight put on the heighest layers ($\sim h^{5/3}$) 
in the $\theta_0$ calculation, causes that mainly variations of the highest layers will drive the spread of the distribution of the 
$\theta_0$ selected TP. The seeing selected profiles reflect the variability of all layers equally, thus explaining the much wider 
statistical distribution of the seeing selected TPs. 

\section{SODAR based vertical turbulence profiles}
\label{sodarbasedprofiles}
\begin{figure}[h]
\caption{High resolution turbulence profiles measured by the SODAR. The markers indicate the MASS and DIMM seeing measured simultaneously 
with the SODAR above the respective sites. The lines show how the seeing measured by the SODARs accumulates onto the MASS seeing, thus providing 
the seeing an observer would see at the corresponding altitude. Note that these data are non representative for the sites. }
\label{sodarprofiles}
\resizebox{0.9 \textwidth}{!}{\rotatebox{270}{\includegraphics{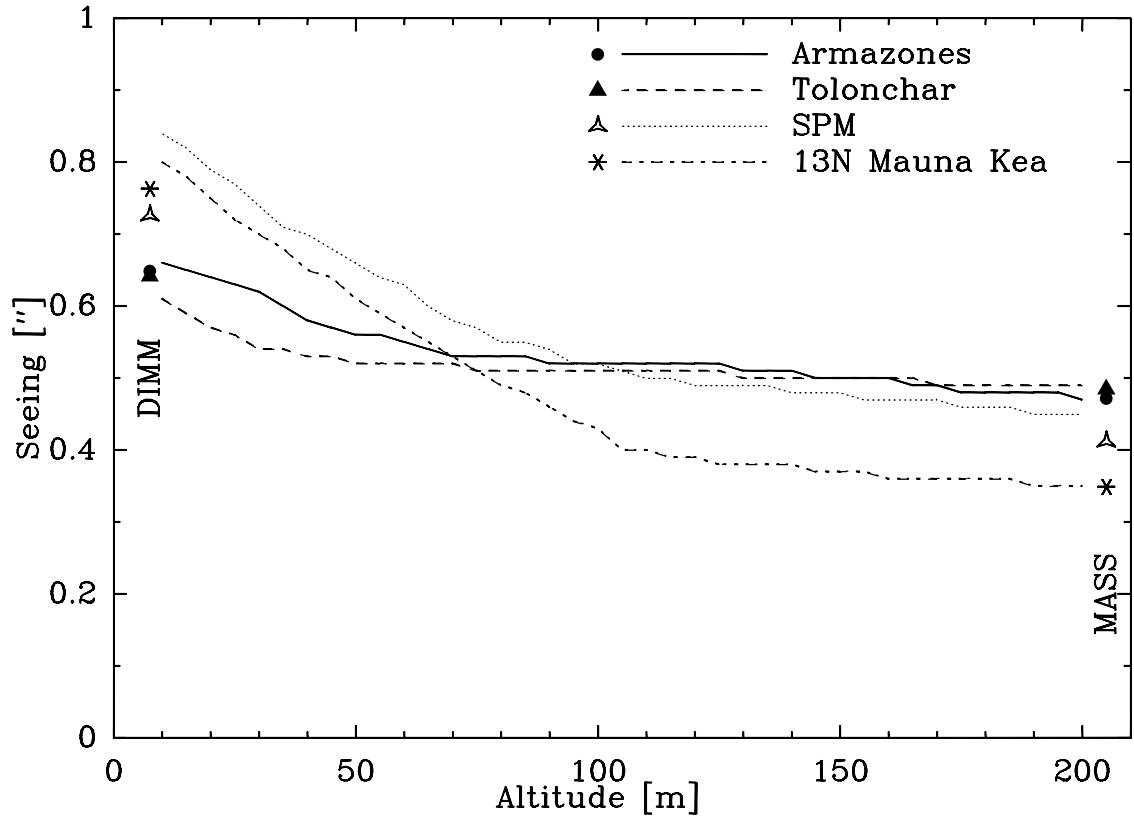}}}
\end{figure}

As the Chilean sites show a weaker GL than the northern sites, but the northern sites show a weaker MASS seeing than the 
Chilean sites, a crossover between the TPs above the northern and southern hemisphere sites has to occur within the lowest 500~m. 
We show now that the SODAR based profiles allow to detect this crossover. In order to compare the SODAR results directly to the 
MASS--DIMM measurements, we selected only MASS-DIMM observations obtained during the 30~minute SODAR data acquisition times. 
For a thorough discussion on the data acquistion, analysis and calibration of the SODAR data we refer to \cite{travouillon08a}. 
Figure~\ref{sodarprofiles} shows the seeing which an observer would see at various altitudes above the ground up to 200~m. 
To obtain this profile, the XFAS profile was integrated from 200~m to 500~m and added to the median MASS seeing value during the time of 
observation. This MASS seeing is indicated by the markers to the right of the SODAR profiles. This 200~m XFAS+MASS seeing was then added to 
the SFAS layers, resulting in the profiles shown in Figure~\ref{sodarprofiles}. The profiles were first added and the median was 
calculated afterwards. The observed DIMM seeing taken simultaneously with the SODAR observations is indicated by the markers to the left of the profiles, 
at 7.5~m above ground. Even though the SODAR profiles start at 10~m above ground, the 10~m datapoint is an integral from 7.5~m to 12.5~m. 
And, as the SODARs were always mounted approximately 1~m to 3~m below the bases of the 6.5~m tall telescope towers, it can be assumed that 
the SODARs do not miss turbulence compared to the DIMM. 

The SODARs were typically operated during much less time than the MASS-DIMM on the sites. These profiles can therefore not be taken as 
representative on absolute scales for the sites. Nevertheless, these profiles allow to draw some interesting conclusions. 

As can be seen from Figure~\ref{sodarprofiles}, the SODAR profiles provide a good match between the MASS and DIMM observations. 
This demonstrates the quality of the data analysis and calibration (\citealt{travouillon08a}). The agreement between SFAS, XFAS, MASS and 
DIMM is better than 10\%, showing that these three methods, each based on a different physical concept, measures the TS with good 
accuracy. The only site where the agreement is worse is SPM. This can be attributed to the increased acoustic noise due to the 
trees surrounding the site, which affect the SODAR measurements (see \citealt{travouillon08a} for details). 

It can also be seen in Figure~\ref{sodarprofiles} that the profiles obtained at the Chilean sites are similar to each other. 
Their TPs show that the bulk of the GL is located below 40~m and 60~m at Tolonchar and Armazones, respectively. 
In comparison, the northern sites start from a weaker TS at higher altitudes, but below 100~m they increase much stronger 
than the Chilean sites. Between approximately 60~m and 90~m, their profiles are crossing the TS of the Chilean sites. 

From the presented turbulence profiles we conclude, that the topography and surface properties of the sites dominate the optical TS
over the lowest hundred meters above ground, which are the main drivers for the overall seeing. The site at SPM is surrounded by trees 
and we believe that the therefore increased surface roughness, heat capacity, and release of humidity, cause the enhanced TS; 
this raises the effective height of the ground. 
As 13N is located approximately 150~m below the summit ridge of Mauna Kea, it falls below the wake region when winds are coming from 
eastern directions, as reported in \cite{skidmore08}. In comparison, the Chilean sites are located on the actual 
summit of the candidate mountains. The terrain shows low surface roughness and is basically free of vegetation. These sites are are 
very dry and all of similar soil properties. These arguments hold specifically for Armazones and Tolar. Cerro Tolonchar is different 
in the sense that it represents a summit plateau with an extension of approximately 500~m in North--South and 250~m in East--West direction
and that some sparse vegetation is present. But as the location of the Tolonchar site testing equipment is on the northern 
most edge of this plateau and winds are predominantely coming from the North-West (\citealt{skidmore08a}), the air above that location 
resembles closely the free air flow.

\section{Seasonal variations of the turbulence strength}
\label{seasonalprofiles}
\begin{figure}[h]
\caption{The standard year of median turbulence strength of each MASS layer.}
\label{txcnyear}
\resizebox{0.9 \textwidth}{!}{\rotatebox{0}{\includegraphics{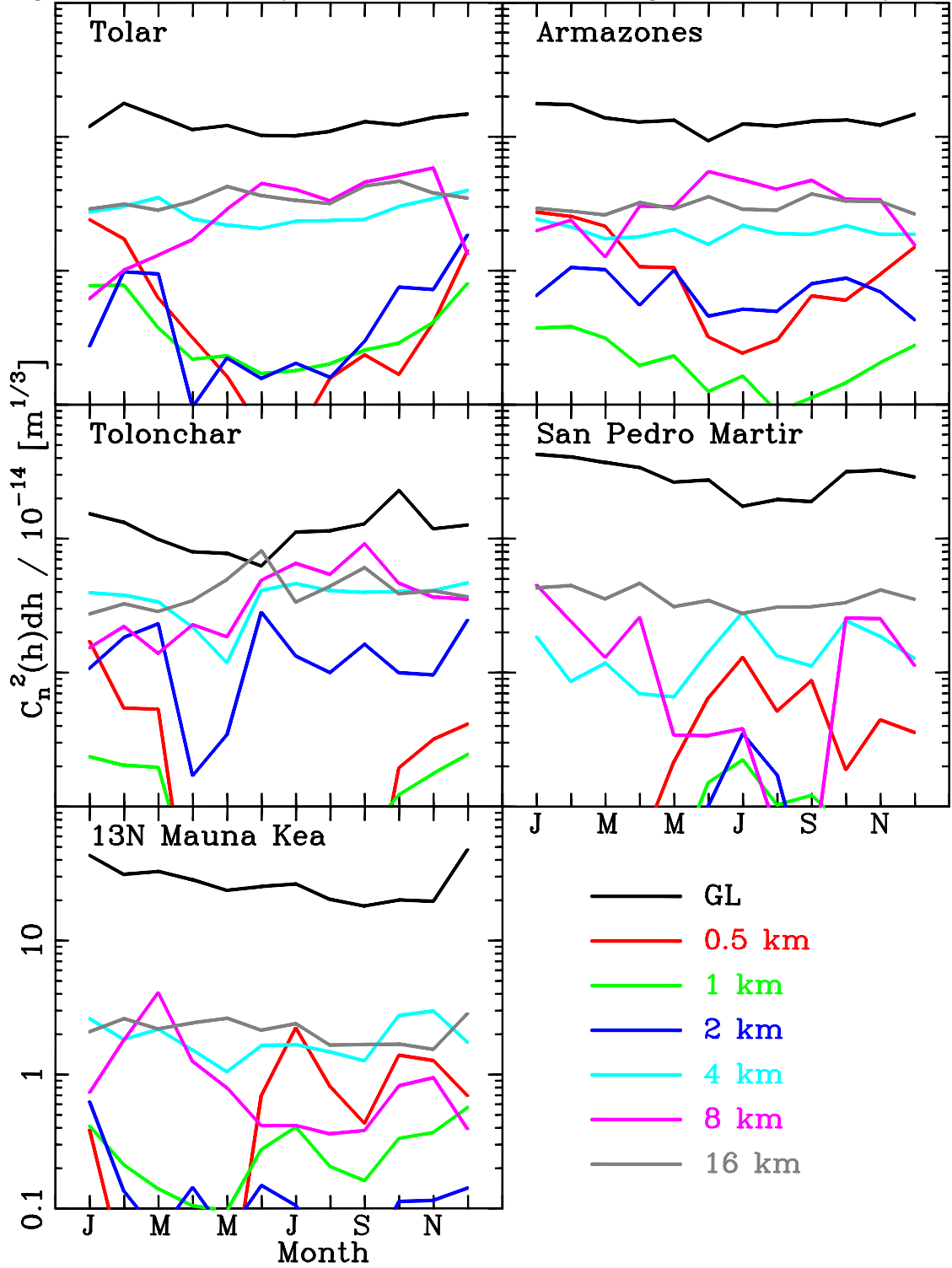}}}
\end{figure}

In recent years, indications for seasonal variability of TS above other observatories were found (\citealt{masciadri06}, \citealt{masciadri01}). 
Clearly revealing the presence of seasonal variability in the TS at various altitudes will help to shed light on the mechanisms driving 
optical turbulence and will be essential for validating numerical models which link meteorological modeling to optical TS. 
To look for seasonal variations of the vertical turbulence distribution throughout the atmosphere in the data we present here, 
we constructed what we call a standard year. For each month of the year, the median TS of each layer was 
computed from all measurements taken during the corresponding months. 
The resulting standard years for the turbulence distribution above the TMT candidate sites are shown in Figure~\ref{txcnyear}. 
Due to technical problems, bad weather and the daylight seasons, not all months contain the same amount of data. To judge the significance 
of features seen in the median TS data, we again refer to Figure~\ref{datapts} which shows how many MASS-DIMM profiles were used for the 
computation of TS in each month of the standard year. Several features in Figure~\ref{txcnyear} are coherent at several sites and in 
various layers to a level of a few $10^{-15}$~m$^{1/3}$. This indicates that the resolution of the MASS is likely 
better than a few $10^{-15}$~m$^{1/3}$.
As can be seen, the seasonal behaviour of the MASS layers below approximately 4~km is anticorrelated with the MASS layers at 8~km. 
The 4~km and 16~km layer exhibit less pronounced seasonal variations at some sites, if any. The 8~km MASS layer  ($\approx$ 10~km to 12~km a.s.l.) 
at the sites falls closest to the altitude domain of the tropopause. The lower MASS layers thus trace the troposphere, whereas the 
16~km MASS layer ($\approx$ 18~km to 20~km a.s.l.) rather traces the lower stratosphere. 
That the parts of the atmosphere below and above this altitude behave differently is therefore not unexpected. We will discuss these 
two altitude regimes separately. 

\subsection{Low altitudes}
First we discuss the Chilean sites. Above Tolar all layers up to 4~km show the lowest TS between May and August, raising to their 
highest values during the summer months. Above Armazones this behaviour is less pronounced and only clearly detectable in the 0.5~km and 1~km 
layer, maybe some weak variation is visible in the 2~km layer as well. Tolonchar shows seasonal trends also only in its 0.5~km and 1~km layer. 
The TS of the 2~km and 4~km MASS layers at Tolonchar show very low TS during April and May. As at Tolonchar only few data are available 
during the months of May and June, the feature in May represents a single year only and it is questionable whether it can be taken as typical for 
this site. Still, there appears to be a maximum altitude of approximately 6~km asl below which seasonal variations -- weak TS in winter and strong 
TS during summer -- can be detected in the atmosphere above northern Chile. 

At these low altitudes, local terrain effects will play an imporant role. Like the proximity of Tolar to the Pacific Ocean which acts as heat reservoir; 
and Tolonchar rising steeply out of its surrounding arid terrain, ranging already into the westerly winds which dominate at higher elevations (\citealt{cuevas09}). 
But also the large scale atmospheric conditions might be the driver of the TS. The ``Altiplano Winter'' phenomenon is a result of the 
shift of the inner tropical convergence zone southwards over the Amazonia region. Over South America this results in increased low altitude winds and convective, 
monsoon-like, activity (\citealt{satyamurty98}, \citealt{zou98}). 
During summer the Chilean sites are located within the region where the parts of these northerly winds which cross the Andes and the 
anti-cyclonic southerlies meet and turn westwards off the Pacific coast. Even though wind speeds in the region of the sites are low, one might 
expect that, as these flows merge, will result in increased large scale wind shear; in addition to the convectivity of the notherlies. 

The situation at the sites of SPM and 13N is somewhat different to what is observed in Chile. The layers up to 2~km, in particular the 0.5~km one, 
are basically turbulence free until May. The 0.5~km layer then becomes strong (up to $10^{-14}$ m$^{1/3}$) from June onwards. This behaviour does not 
show a smooth seasonal variation, like temperature or humidity measured at these sites. Only the 1~km and 2~km layers above 13N show a rather smooth 
seasonal variation on a very low level (few $10^{-15}$~m$^{-1/3}$). 
These variations are opposite to what is observed in Chile: strong TS in winter and weak TS during summer. At SPM 
the seasonal variation is not reaching into the 4~km MASS layer, whereas 13N might show some faint seasonal trend of this layer. 
Interestingly, the behaviour of the 0.5~km layer does not correspond to the GL. For example, during JJA the 0.5~km layer becomes strongest at SPM 
whereas its GL is weakest during these months. 
Both sites are not far to the Pacific Ocean and we expect it to play an important role in the TS. But their local terrain is quite complex 
(vegetation at SPM and 13N being not on the actual summit), not allowing an easy qualitative assessment of the TS behaviour . For the large scale flows it 
can be said that both sites are located in the flow of the Hawaiian high which extends further south during the summer months. During the winter it 
shifts towards north and the north-westerly flow in the area of Hawaii interacts with the south-easterly wind which is driven by the Aleutian low (\citealt{vincent98}). 
This might explain the seasonal variation in the 1~km and 2~km layers at 13N. 

\subsection{High altitudes}

Of the higher altitudes, the 8~km layer shows the stronger seasonal variation, whereas the 16~km layer shows essentialy no seasonal trend. 
The very strong seasonal variation of the 8~km layer, of up to $3\cdot 10^{-14}$~m$^{1/3}$, is the main driver for the seasonal variability of 
the isoplanatic angle at some of the sites (\citealt{skidmore08}). 

\begin{figure}[h]
\caption{Correlation between the standard year turbulence strength of the 8~km MASS layer and NCEP reanalysis MASS weighted wind speed in this MASS layer. 
Only NCEP data were used which fall within $\pm$3~hrs of MD observations. The solid lines show linear fits to the data. }
\label{ncepcn2}
\resizebox{0.9 \textwidth}{!}{\rotatebox{0}{\includegraphics{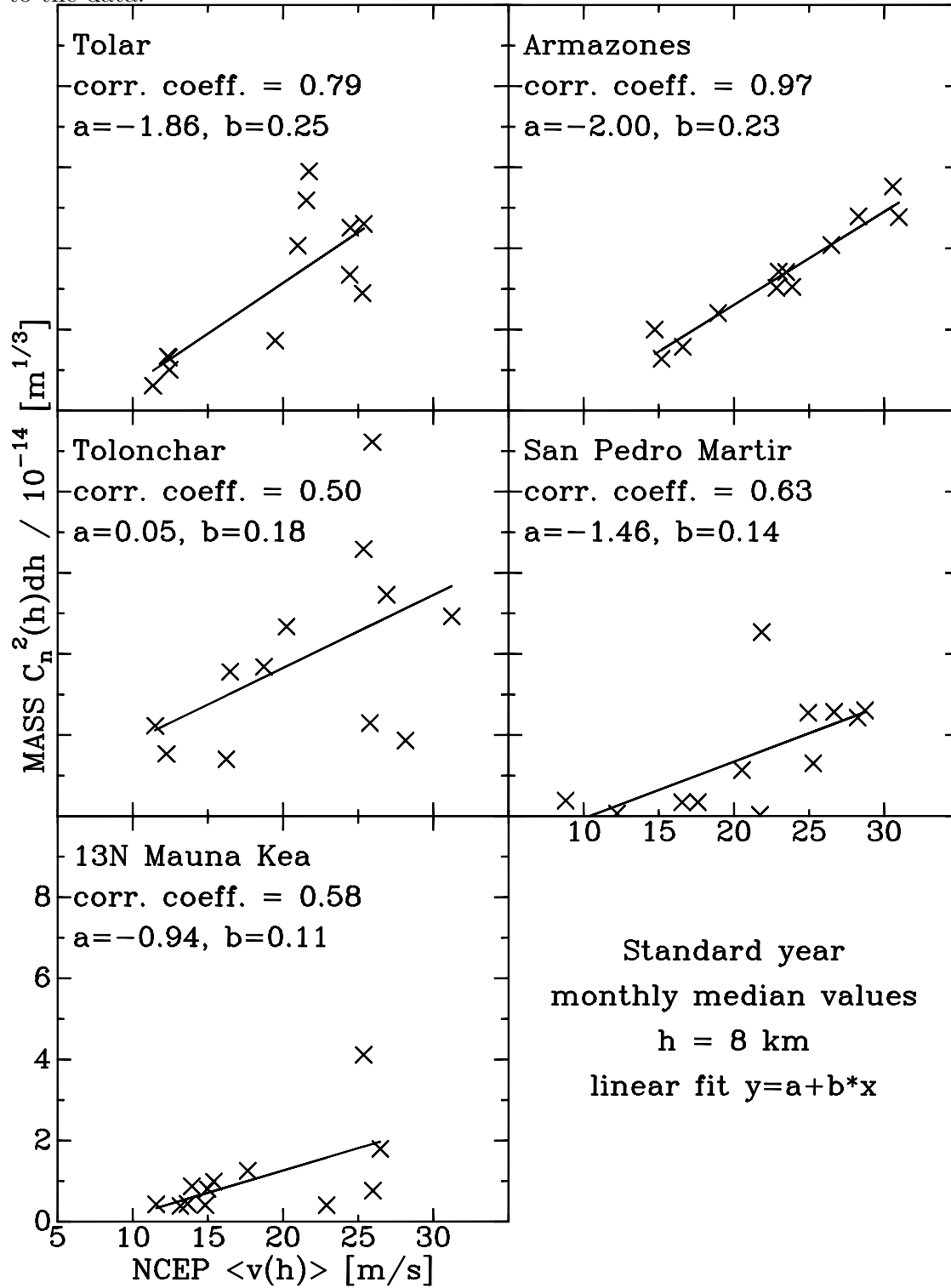}}}
\end{figure}

With altitudes of the TMT candidate sites ranging from 2.3 to 4.5~km asl, the 8~km MASS layer traces 
the domain of the jetstream, whose typical altiude is around 200~mb (corresponding to $\sim$11~km asl). 
An influence of the jetstream activity, or more general of the wind speed, on the TS has been discussed since 
some time in the literature (\citealt{vernin86}, \citealt{sarazin02}, \citealt{masciadri06}). 
To see whether the wind speed is indeed one of the main drivers for the TS at these altitudes, 
we make use of the vertical wind speed profiles provided by the the NCEP/NCAR reanalysis project 
(\citealt{kalnay96}\footnote{NCEP Reanalysis data provided by the NOAA/OAR/ESRL PSD, Boulder, Colorado, USA, from their Web site at {\tt http://www.cdc.noaa.gov/}}).  
The used locations of the NCEP models are given in Table~\ref{ncepcoord}. 
The NCEP wind speed profiles are provided every six hours and at 16 altitudes, ranging from 761~m to 25~km asl. 
Of these, only 13 or 14 levels are above the candidate sites, depending on the sites elevation. 
To correlate these data with the MASS observations, the wind speed profiles have to be convolved with the MASS 
weighting function of the corresponding layer. This was done after interpolating the wind speed profile linearly 
onto a grid with 100~m vertical resolution. Again, we computed a standard year for these wind speed profiles, but only from those data 
for which MASS-DIMM observations were taken within $\pm$3~hours. Figure~\ref{ncepcn2} shows the resulting scatter plots for the 
standard year data of the $h=8$~km layer above  the sites. In particular Armazones shows an excellent linear correlation. 
Tolar and Tolonchar do show a similar increase of the TS with increasing wind speed, however, with more scatter than Armazones. 
In the case of Tolonchar this enhanced scatter is in part due to low statistics during certain months (see above). Nevertheless, 
the slope (increase of TS with wind speed, marked as parameter $b$ in Fig.~\ref{ncepcn2}) is at the same level at all three sites. 
On the other hand, above SPM and 13N the TS remains at very low levels up to $v_{\mathrm{NCEP}}(8~\mathrm{km})dh \approx 20~$m/s 
and only the higher wind speeds show a trend towards stronger $C_n^2(8~\mathrm{km})dh \approx 2\cdot 10^{-14}~$m$^{1/3}$ values. 
In other words, for the same level of wind speed, the turbulence strength at the northern sites is only about half of what 
is observed above Chile. Also the rate of increase is only half of what is observed at the southern sites. We tested similar 
correlations for the other MASS layers but these do not show significant correlations.  

This confirms the findings by \cite{vernin86} who showed a direct correlation between the 200~mbar wind speed and the total seeing observed at 
telescopes on Mauna Kea and La Silla. \cite{vernin86} deduced a threshold of 20~m/s at the 200~mbar level for good total seeing conditions. This is 
remarkably similar to our observations taking into account that \cite{vernin86} had to employ correction factors for the various seeing values and 
could only use the total seeing. \cite{sarazin02} later found from balloon data a correlation of $\tau_0$ with the 200~mbar wind speed above Pach\'on. 
This led to a number of studies of high altitude winds above other existing and potential observatory sites (e.g., \citealt{carrasco05}, \citealt{garcia05}).
It is in principle understandable that high wind speeds are more likely to develop stronger TS, but our observations of the weaker increase 
of TS above the sites of 13N and SPM are at first puzzling and are in contrast to the common threshold found in the \cite{vernin86} data. This could  
indicate that the MASS misses to detect the jetstream above SPM and 13N. But as the MASS calculates the total seeing from the integral of the reconstructed 
$C_n^2(h)dh$ values and also from the total scintillation, it is possible to test the overall integration scheme of the MASS system. We found that both methods 
deliver values of the total seeing which agree to within 10\%: the Armazones median MASS seeing from the scintillation calulation resulted in 0\arcsecf40, 
as compared to 0\arcsecf43 from the $C_n^2$ reconstruction (for 13N Mauna Kea the numbers read 0\arcsecf30 and 0\arcsecf34, respectively). 
We believe that this, in combination with the previously shown agreement with other instruments, demonstrates that no turbulence is acutally missed by the MASS. 
On the other hand, the interpolated wind speed will certainly not be very accurate as the original NCEP altitude grid is not dense. We experimented with different 
interpolation schemes, as well as using the raw wind speed from the closest level to 8~km above site altitude and all resulted in very similar figures.

The observed differences between the correlations would then show that the 200~mb wind speed alone is not a good indicator on a global scale for the median TS at 
these altitudes and that the jetstream above the northern sites during the times of our observations was somewhat less turbulent than above the Chilean mountains. 
This might be caused by the lack of a large continental land mass and a high mountain range, as in the case of the Andes mountains. In northern Chile, the 
airflow over the South American continent, which rises steeply out of the sea, can be expected to show enhanced formation of atmospheric waves.  
Such waves can also be forced by differential heating patterns like the proximity of the Atacama desert to the Pacific Ocean. 
Atmospheric waves are commonly thought to result in an increased TS. Unfortunately, such waves and their turbulence production cannot yet be modeled properly, 
even though comparisons between models and observations show promising results (e.g., \citealt{jumper07}). Our results might indicate that it is the actual 
interaction between wave and jet stream which is driving the TS.  

\section{Night time variations of the turbulence strength}
\label{nightprofiles}
\begin{figure}[h]
\caption{Behaviour of median turbulence strength during each hour after sunset. To avoid the seasonal 
length of the nights affecting the data, this plot shows from 5~hours onwards the turbulence strength 
calculated before sun rise. }
\label{txcnnight}
\resizebox{0.9 \textwidth}{!}{\rotatebox{0}{\includegraphics{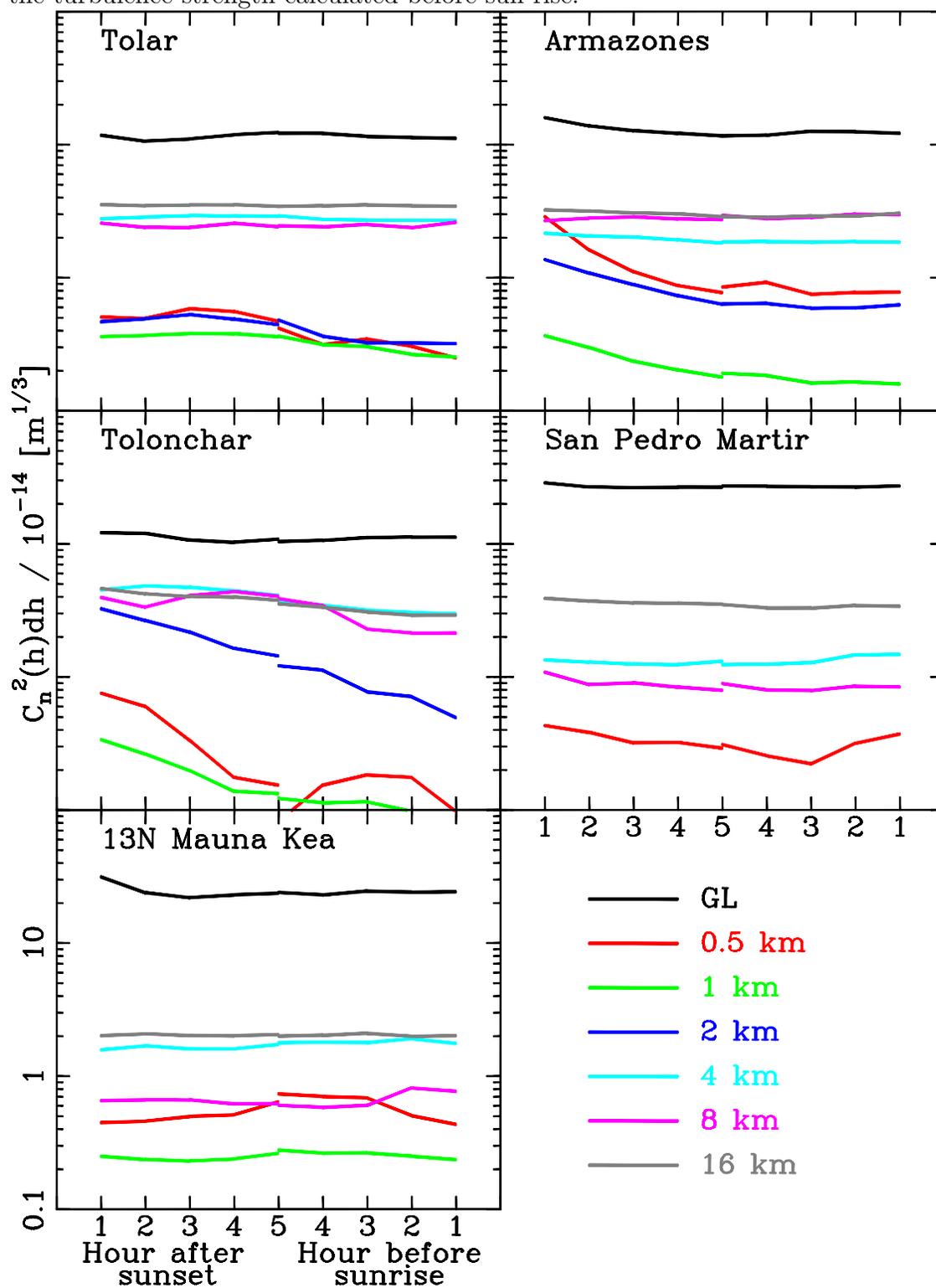}}}
\end{figure}

To investigate the typical behaviour of the TS during the night we computed, similar to the standard year, a standard night. It consists of the median 
TS of all data collected during each hour after sunset. Figure~\ref{txcnnight} shows the results for the sites. To avoid artefacts introduced by the
seasonal variations of the times of sunrise and sunset, the graphs in Figure~\ref{txcnnight} represent the merged behaviour for the hourly medians computed with respect 
to sunset and sunrise. We found that the curves computed with respect to sunrise and sunset match best when midnight is set to 5~hours and the standard night 
for which data are available, is therefore 11~hours long. The jumps of the curves at 5~hours in Figure~\ref{txcnnight} are the offset between the curves computed 
with respect to sunset and sunrise. 

Again, it can be noted that coherent structures are apparent in these graphs on a scale of a few $10^{-15}$~m$^{1/3}$,  
consistent with what was found in Section~\ref{seasonalprofiles}. Also, the median night time evolution of the TS indicates that the atmosphere 
above approximately 4~km above the sites is basically decoupled from air below. The high altitude layers remain close to constant throughout the night, whereas 
the lower layers show a decreasing TS with time. Only 13N and Tolar exhibit almost no change during the course of the night in the lower layers. 
At SPM the 0.5~km layers at this site shows a decrease of its TS until one two hours before sunrise. 
The TS in the layers up to 2~km above Armazones decrease during the first half of the night and then remain at almost constant values. Finally, the 
air above Tolonchar shows a continuous drop of TS throught the night, which maybe only levels off in the 0.5~km layer. 

\begin{figure}[h]
\caption{Behaviour of median wind speed (upper panel) and air temperature (lower panel) at 2~m above ground during each hour after sunset at the 
candidate sites (similar to Fig.~\ref{txcnnight}). }
\label{mednightwst}
\resizebox{0.75 \textwidth}{!}{\rotatebox{270}{\includegraphics{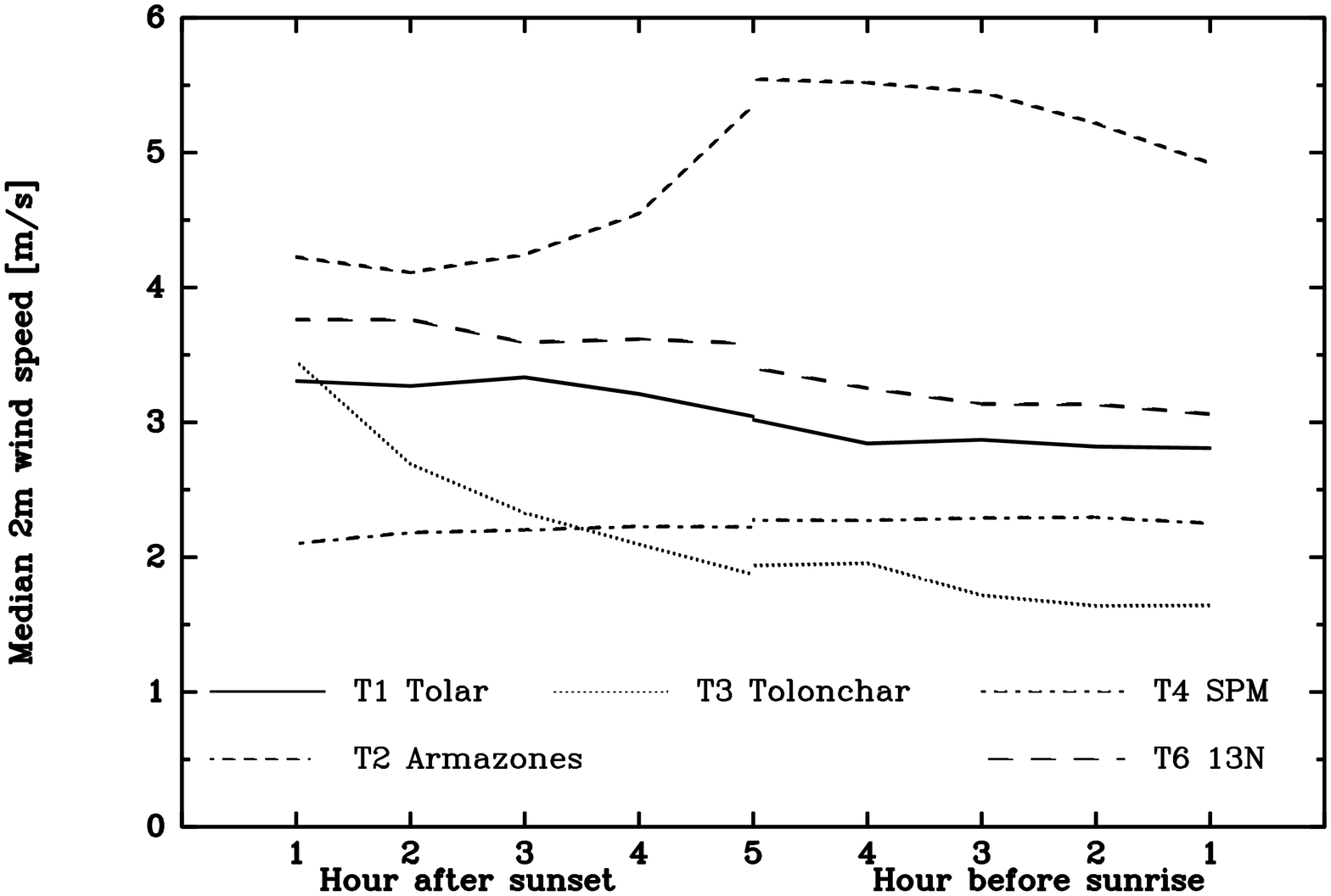}}}

\resizebox{0.75 \textwidth}{!}{\rotatebox{270}{\includegraphics{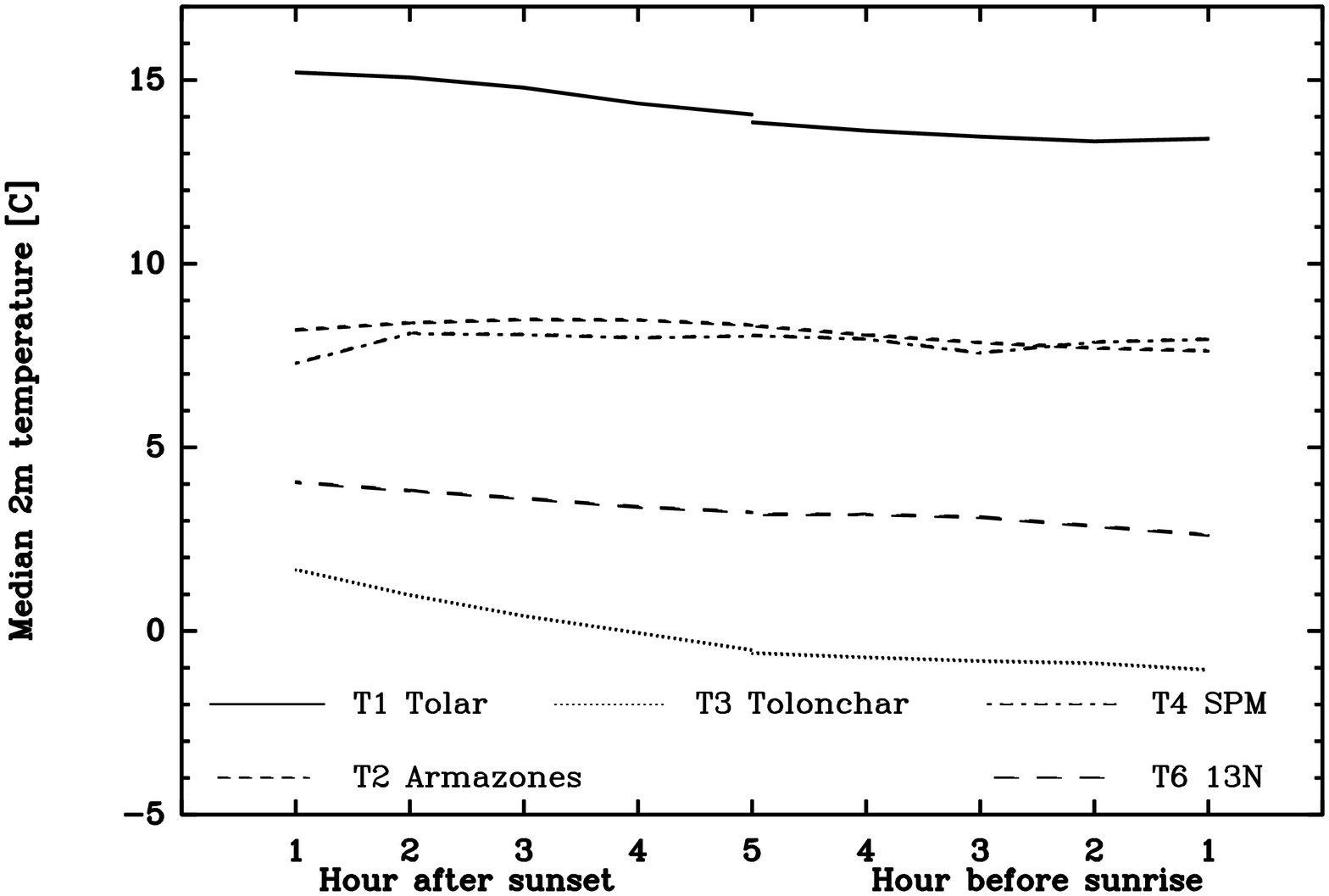}}}
\end{figure}

The observed behaviour appears similar to a cooling process and the differences between the sites on how the TS evolves below the tropopause, 
are related to the evolution of wind shear and temperature gradients. We do not have measurements in order to compute the Richardson number or any 
other stability parameter in this altitude range. Nevertheless, we make the assumption that the temperature and wind speed measured at the sites are related 
to their respective gradients throughout the air above the site. This assumption is only strictly true below the height of the boundary layer which,  
based on the SODAR profiles presented earlier, we expect to be in the order of 100~m. 
Figure~\ref{mednightwst} shows the standard night for the wind speed and temperature, recorded 2~m above ground simultaneous to the TP measurements. 
It can be seen that the temperature descreases, similarly at all sites. In contrast, the nightly evolution of the wind speed shows remarkable differences 
between the sites. In particular Armazones shows an increase during the second half of the night, whereas Tolonchar shows a continuous decrease of 
the measured wind speed. Similar to Tolonchar, but at a lower level, is the behaviour at 13N and Tolar. Note that our wind speed measurements are unreliable 
at SPM due to the site being surrounded by trees. Therefore, we will not try to interpret the behaviour at SPM. 

From these observations we suspect that cooling of the terrain during the night is important for the decrease of the TS. This is in particular the case 
for continental sites such as Tolonchar and Armazones. We attribute the constant TS at Armazones during the second half of the night to increased 
wind shear. Due to its much higher heat capacity and its better efficiency to carry thermal energy below its surface (\citealt{stull88}), the sea surface temperature of the Pacific Ocean  
remains almost constant during night time. This certainly affects the turbulent fluxes within the air above and close to the ocean. 
We therefore suspect the proximity of the sites of Tolar and 13N to the Pacific Ocean as reason for the almost constant TS observed at these sites. 
SPM being located in the center of the peninsula Baja California appears to be rather an inland case. 

\section{Turbulence profiles at different wind speeds}
\label{meteorelations}

The wind speed plays a crucial role as turbulence driver by introducing shear in the flow. 
Here the turbulence measurements of each MASS layer are correlated to the wind speed which was measured 2~m above 
ground by the automated weather stations. 
These weather stations were equipped with cup anemometers of type AN3 by Monitor Sensors. 
In principle we could have also used the wind measurements taken by the CSAT-3 sonic anemometers which were installed close to the telescopes 
at 7~m above ground (\citealt{riddle08}). The advantage of sonic anemometer measurements is that mechanical anemometers are prone to problems induced by dust 
or ice in the mechanisms. However, we found the two systems to agree very well during most of the time. A particular case is Tolonchar, where 
a strong gradient exists between the 2~m and 7~m level, due to the steep terrain drop towards the prevailing wind direction.  
Still, we show here only the AWS anemometer data, as the sonic anemometers were installed only during the second half of the site testing program 
and thus do not cover the full time span of MASS-DIMM measurements. 

\begin{figure}[h]
\caption{Median turbulence strength of each MASS layer as a function of the wind speed. The median TS 
was computed for 1~m/s wind speed bins. }
\label{cn2vsws}
\resizebox{0.9 \textwidth}{!}{\rotatebox{0}{\includegraphics{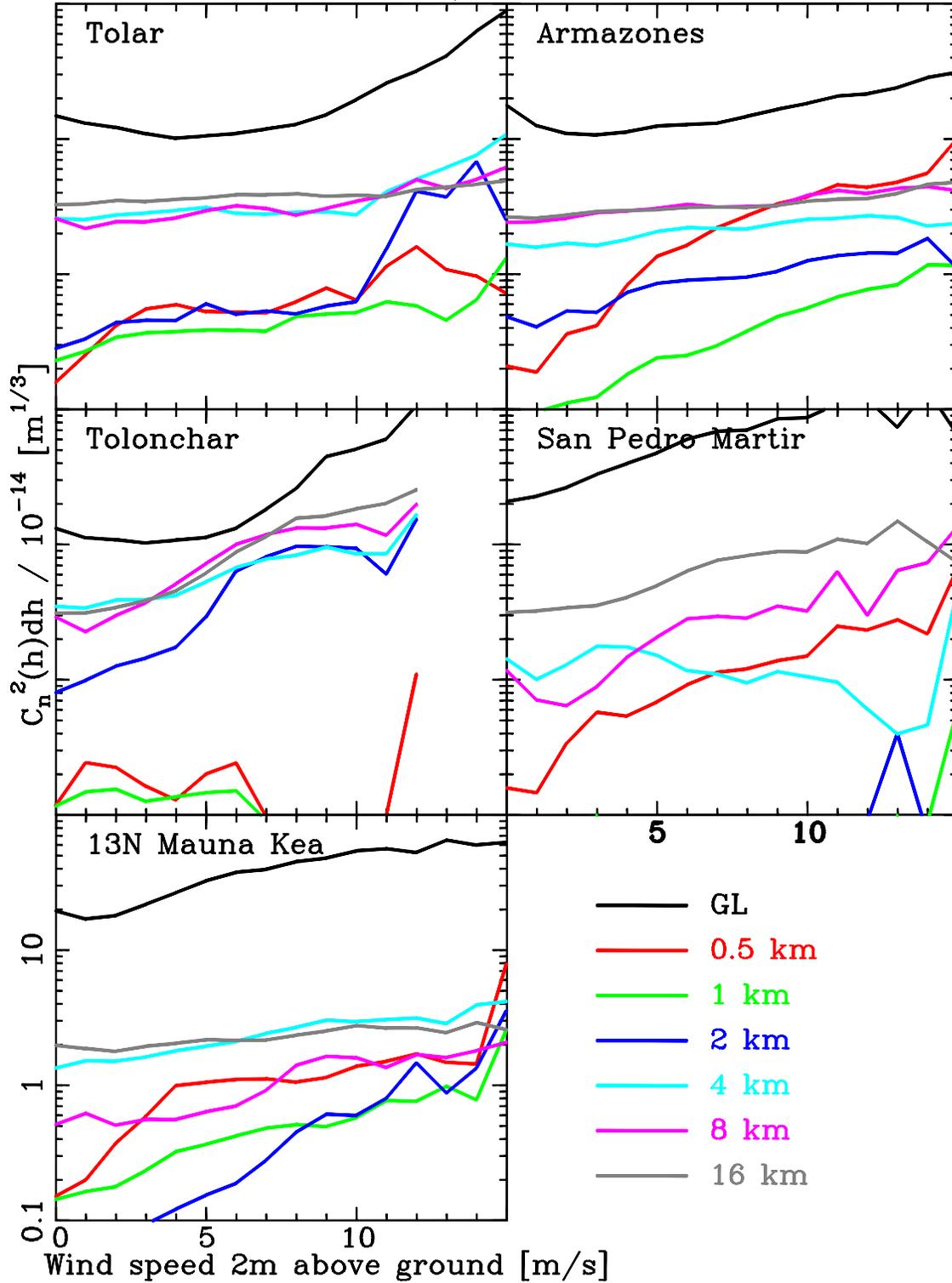}}}
\end{figure}

\begin{figure}[h]
\caption{Correlation coefficient between NCEP wind speed at various altitudes and the 
measured wind speed by the AWS at the sites (see text). Amount of data points is indicated in brackets. 
Altitudes of the sites marked by arrows. }
\label{awsncepcorr}
\resizebox{0.9 \textwidth}{!}{\rotatebox{270}{\includegraphics{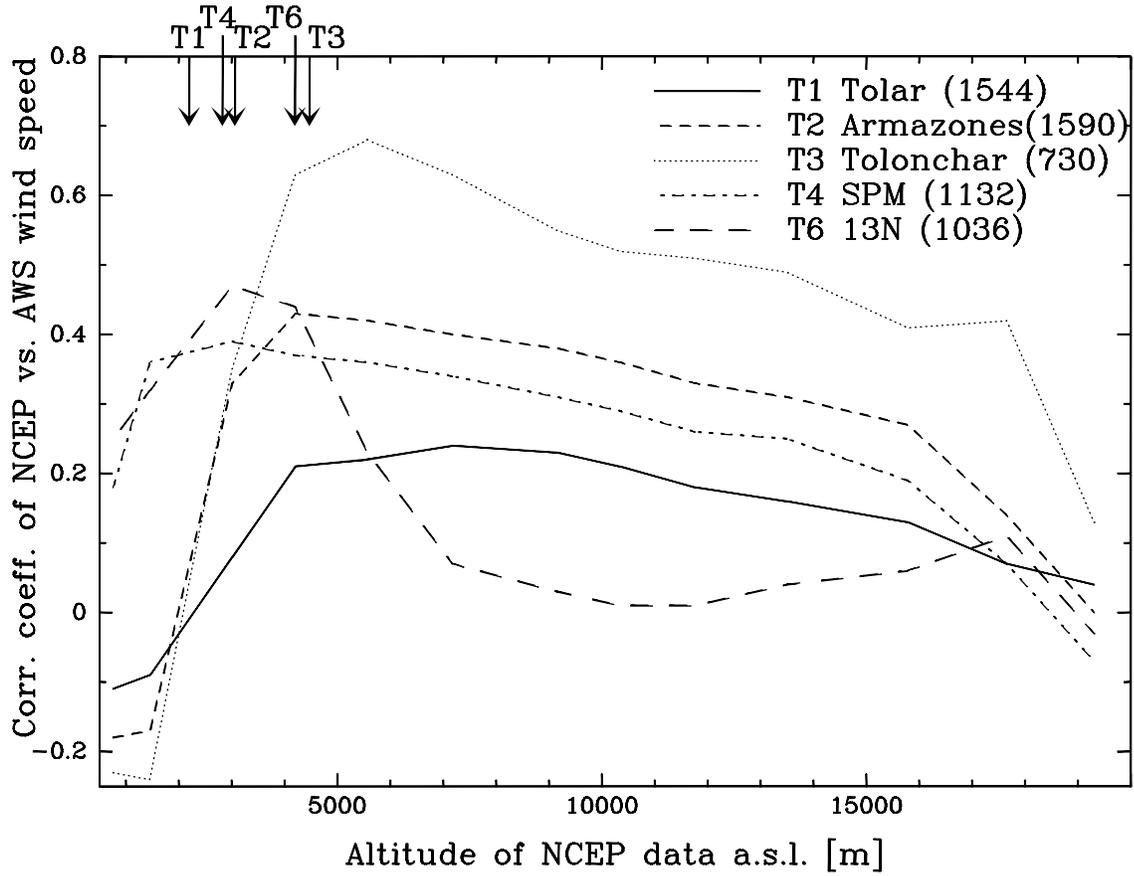}}}
\end{figure}

The panels of Figure~\ref{cn2vsws} show the dependence of the TSs of the individual MASS layers on the wind speed. The median $C_n^2(h)dh$ value 
was computed in 1~m/s wind speed bins for wind speeds up to the operational limit of the site testing telescope. This limit was typically 15~m/s, only 
the Tolonchar setup was limited to 12~m/s due to the strong gradient mentioned before. We expect that the measured wind speed at 2~m above ground is 
indicative of the wind speed within the lowest MASS layers. For the higher layers this assumption can not be expected to hold, as the wind adjusts 
here to the geostrophic conditions which differ from the situation close to the ground. Nevertheless, some correlation might still be present. 
To visualize this behaviour Figure~\ref{awsncepcorr} shows the correlation coefficient of each NCEP model wind speed with the measured wind speed 2~m 
above the sites. Again, measured wind speeds were simultaneous to MASS-DIMM observations and the median was calculated for data taken within $\pm$3 hours 
around the NCEP model times. The strongest correlations, which is always positive, are found at altitudes slightly above the sites. 
The correlations break down at altitudes below the sites. 

As was seen in the overall profiles, as well as in \cite{skidmore08}, the ground layer at all sites shows the strongest TS by up 
to a factor of a few. This is not unexpected, as the MASS-DIMM defined ground layer covers (most of) the atmospheric 
boundary layer, i.e., the layer in which turbulence production is dominated by the interaction between air and ground. Shearing 
forces are strong in this layer as the air speed at surface has to reach zero. Also the ground might heat or cool the air 
above it by means of the sensible heat flux, introducing temperature gradients. As was already shown in \cite{skidmore08}, the GL shows the 
lowest $C_n^2dh$ in the range of wind speeds between approximately 2~m/s and 8~m/s for the Chilean sites and approximately 2~m/s and 3~m/s for 
13N Mauna Kea. SPM shows a continuous increase of $C_n^2dh$ of the GL with increasing wind speed. As that 
site is surrounded by trees the wind speed measurements, especially at low values, are doubtful. Nevertheless, the observed 
increase of the GL with wind speed appears reasonable also at that site. 

The behaviour of the 0.5, 1 and 2~km layers above the sites seems to be dominated by the orography of the surrounding terrain. 
These layers above Tolar remain at the same level of TS up to 10~m/s. At Armazones the 0.5~km layer rises strongly, whereas its
1 and 2~km layers rise less steep. These trends are similar to what happens above 13N. Above Tolonchar the 0.5~km and 1~km layers 
remain at very low TS, but the 2~km layer rises up to 5~m/s and then shows a quite strong increase by $2\cdot 10^{14}$~m$^{1/3}$ towards 7~m/s. 
Tolonchar is about 2000m above the level of the Salar de Atacama (a flat dry salt lake), which is in the prevailing wind direction and within 20~km from the mountain base.
These 20~km are smoothly sloped terrain ranging up 500m-800m below the Tolonchar summit. Therefore, one could assume that the wind reaching the summit 
effectively resembles an altitude of 1000-2000m above the surface and the 0.5 and 1~km MASS layers trace the free air flow between 2 and 3~km above the 
effective terrain. Tolar and Armazones reach out similarly but not as high and steep as Tolonchar. In the case of Armazones a mountain range which is slightly 
higher (3400~m asl) is located in the prevailing wind direction. We speculate that these orographic features cause the onset of waves, driving the TS at 
these altitudes. This picture seems reasonable also for the 13N site, as it is below the actual summit. As our SPM wind speed data are not reliable due to 
the presence of trees, we rather do not want to fit it into this picture. Nevertheless, the increase in TS of the 0.5~km layer appears reasonable. 

The 8 and 16~km layers show as well an increase in their TS with wind speed at the sites, in particular Tolonchar. From Figure~\ref{awsncepcorr} it turns 
out that Tolonchar shows the highest correlation of all sites between the measured and modeled wind speeds. This indicates that its altitude brings it 
into the regime of the geostrophic westerly flows which are probably better modelled by NCEP and are less dependent on the local orography and 
therefore the correlation is higher. 

\section{Summary and conclusions}
The MASS and SODAR based turbulence profiles collected at the TMT candidate sites have been presented. 
The average profiles explain the origin of the observed behaviour of the integrated turbulence parameters. 
At all sites the turbulence strength of the lowest 500~m -- the ground layer -- is found to be the dominating contributor to the overall seeing. 
This layer is stronger at 13N and SPM as compared to the Chilean sites. On the other hand, 13N and SPM show lower turbulence strengths at higher altitudes, 
resulting in a weaker MASS seeing than at the Chilean sites. 
The SODAR measurements, even though these cover much less time than the MASS observations, support this finding. 
They indicate that the altitude dependence of the overall seeing at 13N and SPM is steeper than at the Chilean sites between approximately 60~m and 100~m 
above ground. This results in a stronger total seeing below this altitude at the northern sites as compared to the Chilean sites.
The agreement found between the measurements of MASS, DIMM and SODAR of 10\% demonstrates that these techniques do all have a good accuracy
to measure $C_n^2$. However, we expect the SODAR measurements to be less accurate (\citealt{travouillon08a}) than the MASS-DIMM measurements. 
The isoplanatic angle differences between the sites depends on the distribution of the high altitude turbulence. In this respect, SPM shows similar 
high altitude turbulence strengths as the Chilean sites and therefore a similar isoplanatic angle, despite its MASS seeing being comparable to 13N. 

Seasonal and nocturnal variations of the median turbulence strength recorded by the MASS have also been shown. These data indicate that the resolution 
of the MASS is in the order of a few $10^{-15}$~m$^{1/3}$. Clear seasonal variations are observed up to 2~km and at 8~km above the sites. The 4~km and 16~km layers
show only little or no seasonal trends. The 8~km MASS layer turbulence strength was found to correlate well with the wind speeds at these altitudes above 
the Chilean sites. As seasonal turbulence strengths of the 8~km and the lower MASS layers are basically anit correlated, it can be expected that ground layer 
adaptive optics will be most efficient during the summer months at all sites. 
Nocturnal variations are typically confined to regions below the 4~km MASS layer. 
These trends seem to indicate that orographic features and surface energy properties on scales of tens of km are important drivers for 
the turbulence strength at altitudes below the tropopause. In particular the close proximity of the Pacific Ocean to some of the sites is dominating 
the behaviour of the turbulence strengths at altitudes above approximately 200~m. 

Astronomical turbulence profilers provide a powerful tool to sense clear sky atmopsheric turbulence. The collected data 
are thus not only providing the necessary information for the decision at which site to build TMT and for the design of 
future astronomical instruments. These data have also to be analysed in view of their relation to meteorological mechanisms. The present study could 
only provide a brief overview of some aspects of what can be extracted from such data. More detailed studies using climatological and meteorological 
data, computational models in combination with new observations from turbulence profilers should result in a better understanding of the actual drivers 
of optical turbulence in earths atmosphere. 

\acknowledgments
The TMT Project gratefully acknowledges the support of the 
TMT partner institutions. They are the Association of Canadian 
Universities for Research in Astronomy (ACURA), the California 
Institute of Technology and the University of California. 
This work was supported as well by the Gordon and Betty Moore 
Foundation, the Canada Foundation for Innovation, the Ontario 
Ministry of Research and Innovation, the National Research 
Council of Canada, the Natural Sciences and Engineering Research 
Council of Canada, the British Columbia Knowledge Development Fund, 
the Association of Universities for Research in Astronomy (AURA) 
and the U.S. National Science Foundation.


\clearpage
\clearpage
\clearpage
\clearpage
\clearpage
\clearpage
\clearpage
\clearpage
\clearpage
\clearpage
\clearpage
\begin{table}
\begin{center}
\caption{General togopgraphic parameters of the TMT candidate sites and times during which MASS, DIMM and SODAR 
data were collected on the sites.}
\label{sites}
\begin{footnotesize}
\begin{tabular}{crrrrrrrrrrr}
\tableline\tableline
Site name	&	Latitude	& Longitude	&	Elevation		& \multicolumn{2}{c}{Dates between which data are available} \\
and telescope\#	&	[deg N]		&	[deg W]	&	[m a.s.l.]	&  MASS-DIMM  & SODAR \\
\tableline
Cerro Tolar	-- T1	  &	-21.9639	&	70.0997  &	2290	&  Oct. 2003 -- Apr. 2007	&    \\
Cerro Armazones -- T2 & -24.5800	&	70.1833  &	3064	&  Nov. 2004 -- Feb. 2008	& Sep. 2005 -- Dec. 2005 \\
Cerro Tolonchar	-- T3 & -23.9333	&	67.9750	 &	4480	&  Jan. 2006 -- Feb. 2008	& Mar. 2007 -- Feb. 2008 \\
SPM -- T4			  & 31.0456		&	115.4691 &	2830	&  Oct. 2004 -- Feb. 2008	& May  2006 -- Jun. 2006 \\
13N Mauna Kea	-- T6 &  19.8330	&	155.4810 &	4050	&  Jun. 2005 -- Feb. 2008	& Oct. 2005 -- Dec. 2007 \\
\tableline
\end{tabular}
\end{footnotesize}
\end{center}
\end{table}
\clearpage
\begin{table}
\begin{center}
\caption{Used NCEP/NCAR reanalysis locations for the TMT candidate sites.}
\label{ncepcoord}
\begin{tabular}{crrrrrrrrrrr}
\tableline\tableline
Site name	&	Latitude	& Longitude	  \\
and telescope\#	&	[deg N]		&	[deg W] \\
\tableline
Cerro Tolar	-- T1 &	-21.25	&	71.25     \\
Cerro Armazones -- T2 & -23.75	&	71.25  \\
Cerro Tolonchar	-- T3 & -23.75	&	68.75  \\
SPM -- T4 & 31.25	&	116.25   \\
13N Mauna Kea	-- T6 &  18.75	&	156.25  \\
\tableline
\end{tabular}
\end{center}
\end{table}
\clearpage
\appendix
\section{Tabulated turbulence profiles}
\begin{table}
\begin{center}
{\scriptsize
\caption{ Vertical turbulence profiles. Statistics were computed over all data collected for each layer
independently. }
\label{profilestable}
\begin{tabular}{c c c c c c c}
\hline
\hline
height & \multicolumn{6}{c}{$C_n^2~dh ~[m^{1/3}]$}  \\
$[$km$]$ &       25\%ile   &  median  &   75\%ile  & & mean   & rms   \\
\hline
\multicolumn{7}{c}{Cerro Tolar: 196812 total profiles, October 2003 -- April 2007} \\
\hline
 0.0 &           5.26e-14 &           1.11e-13 &           2.02e-13 & &           1.62e-13 &      2.88e-13     \\
 0.5 &           4.21e-18 &           3.84e-15 &           3.09e-14 & &           4.26e-14 &      1.37e-13     \\
 1.0 &           3.52e-16 &           3.21e-15 &           1.37e-14 & &           2.74e-14 &      1.02e-13      \\
 2.0 &           1.78e-16 &           3.88e-15 &           3.49e-14 & &           3.77e-14 &      8.69e-14      \\
 4.0 &           9.80e-15 &           2.77e-14 &           5.52e-14 & &           4.42e-14 &      5.79e-14      \\
 8.0 &           4.09e-15 &           2.54e-14 &           7.72e-14 & &           5.05e-14 &      6.21e-14      \\
16.0 &           2.29e-14 &           3.51e-14 &           5.49e-14 & &           4.57e-14 &      3.96e-14      \\
\hline
\multicolumn{7}{c}{Cerro Armazones: 212367 total profiles, November 2004 -- February 2008} \\
\hline
 0.0 &           6.04e-14 &           1.19e-13 &           2.15e-13 & &           1.71e-13  &     2.77e-13       \\
 0.5 &           1.23e-17 &           9.78e-15 &           5.58e-14 & &           6.93e-14  &     1.85e-13       \\
 1.0 &           6.59e-17 &           1.99e-15 &           1.10e-14 & &           2.38e-14  &     9.17e-14       \\
 2.0 &           5.27e-16 &           7.16e-15 &           3.23e-14 & &           3.28e-14  &     7.44e-14      \\
 4.0 &           6.34e-15 &           1.93e-14 &           4.09e-14 & &           3.33e-14  &     4.66e-14       \\
 8.0 &           8.66e-15 &           3.03e-14 &           7.96e-14 & &           5.63e-14  &     6.87e-14      \\
16.0 &           1.96e-14 &           3.04e-14 &           4.84e-14 & &           4.05e-14  &     3.68e-14      \\
\hline
\multicolumn{7}{c}{Cerro Tolonchar: 89958 total profiles, January 2006 -- February 2008} \\
\hline
 0.0 &           4.72e-14 &           1.04e-13 &           1.69e-13 &  &          1.14e-13  &     1.93e-13       \\
 0.5 &           2.11e-21 &           1.76e-15 &           2.60e-14 & &           3.71e-14  &     1.21e-13      \\
 1.0 &           2.87e-17 &           1.29e-15 &           8.27e-15 & &           2.38e-14  &     9.18e-14      \\
 2.0 &           9.02e-16 &           1.29e-14 &           6.21e-14 & &           6.18e-14  &     1.30e-13      \\
 4.0 &           1.65e-14 &           3.87e-14 &           7.59e-14 & &           6.06e-14  &     7.11e-14      \\
 8.0 &           6.75e-15 &           3.47e-14 &           1.03e-13 & &           6.49e-14  &     7.41e-14      \\
16.0 &           2.24e-14 &           3.62e-14 &           6.12e-14 & &           5.77e-14  &     6.96e-14      \\
\hline
\multicolumn{7}{c}{SPM: 139359 total profiles, October 2004 -- February 2008} \\
\hline
 0.0 &           1.75e-13 &           2.76e-13 &           5.09e-13 & &           4.88e-13  &    7.05e-13       \\
 0.5 &           1.67e-21 &           3.10e-15 &           3.62e-14 & &           6.39e-14  &    2.66e-13          \\
 1.0 &           1.56e-21 &           5.39e-16 &           5.93e-15  & &          2.22e-14  &    1.09e-13       \\
 2.0 &           9.82e-19 &           4.30e-16 &           9.41e-15 & &           2.55e-14  &    8.61e-14         \\
 4.0 &           8.00e-16 &           1.37e-14 &           4.27e-14 & &           3.55e-14  &    5.86e-14         \\
 8.0 &           2.70e-17 &           1.06e-14 &           5.57e-14 &  &          4.15e-14  &    6.57e-14         \\
16.0 &           2.25e-14 &           3.60e-14 &           6.10e-14 & &           5.43e-14  &    6.00e-14        \\
\hline
\multicolumn{7}{c}{13N Mauna Kea: 121801 total profiles, June 2005 -- February 2008} \\
\hline
 0.0 &           1.38e-13 &           2.41e-13 &           4.29e-13 & &           3.63e-13  &  5.07e-13          \\
 0.5 &           1.22e-20 &           5.11e-15 &           4.21e-14 & &           7.71e-14  &  3.11e-13          \\
 1.0 &           1.88e-16 &           2.46e-15 &           1.13e-14 & &           4.03e-14  &  1.83e-13           \\
 2.0 &           3.26e-17 &           8.98e-16 &           1.18e-14 & &           2.55e-14  &  8.15e-14         \\
 4.0 &           4.81e-15 &           1.72e-14 &           4.12e-14 & &           3.48e-14  &  5.56e-14         \\
 8.0 &           4.34e-16 &           6.21e-15 &           3.35e-14 & &           3.07e-14  &  5.58e-14          \\
16.0 &           1.27e-14 &           2.03e-14 &           3.48e-14 & &           3.02e-14  &  3.53e-14          \\
\hline
\end{tabular}
}
\end{center}
\end{table}
\clearpage
\begin{table}
\begin{center}
{\scriptsize
\caption{ Mean and median $C_n^2~dh$ profiles constructed from 10\% of all profiles collected at each site,  
around the 25, 50 and 75 percentiles of the DIMM $r_0$. The number of profiles used and time span covered by 
these are given for each site.  }
\label{profileDIMMseltable}
\begin{tabular}{c c c c c c c c c}
\hline
\hline
height   & \multicolumn{3}{c}{median $C_n^2~dh ~[m^{1/3}]$}      & & \multicolumn{3}{c}{mean $C_n^2~dh ~[m^{1/3}] $} \\
$[$km$]$   &   25\%  $r_0$   &   50\% $r_0$               &  75\% $r_0$  & &   25\%  $r_0$   &  50\% $r_0$     &   75\% $r_0$ \\
\hline
\multicolumn{8}{c}{Cerro Tolar: 19657 used profiles, October 2003 -- April 2007} \\
\hline
 0.0 &     9.41e-14 &     1.31e-13 &     1.81e-13 & &     8.72e-14 &     1.20e-13 &     1.68e-13 \\
 0.5 &     1.17e-15 &     4.29e-15 &     9.85e-15 & &     1.24e-14 &     2.35e-14 &     4.62e-14 \\
 1.0 &     2.14e-15 &     3.53e-15 &     6.03e-15 & &     6.20e-15 &     1.24e-14 &     2.80e-14 \\
 2.0 &     9.79e-16 &     5.26e-15 &     2.08e-14 & &     8.15e-15 &     2.14e-14 &     4.98e-14 \\
 4.0 &     2.04e-14 &     2.99e-14 &     4.51e-14 & &     2.54e-14 &     3.69e-14 &     6.19e-14 \\
 8.0 &     2.09e-14 &     3.15e-14 &     4.24e-14 & &     3.76e-14 &     5.27e-14 &     6.73e-14 \\
16.0 &     3.32e-14 &     3.66e-14 &     3.96e-14 & &     3.90e-14 &     4.55e-14 &     5.33e-14 \\
\hline
\multicolumn{8}{c}{Cerro Armazones: 21238 used profiles, November 2004 -- February 2008} \\
\hline
 0.0 &     1.02e-13 &     1.37e-13 &     2.05e-13 & &     9.64e-14 &     1.29e-13 &     1.91e-13 \\
 0.5 &     2.03e-15 &     1.02e-14 &     3.71e-14 & &     1.20e-14 &     2.94e-14 &     7.86e-14 \\
 1.0 &     9.61e-16 &     2.17e-15 &     5.49e-15 & &     4.29e-15 &     9.47e-15 &     2.62e-14 \\
 2.0 &     2.90e-15 &     1.04e-14 &     1.99e-14 & &     1.01e-14 &     2.42e-14 &     4.54e-14 \\
 4.0 &     1.41e-14 &     2.14e-14 &     3.24e-14 & &     1.84e-14 &     2.75e-14 &     4.69e-14 \\
 8.0 &     2.49e-14 &     3.69e-14 &     4.52e-14 & &     3.89e-14 &     5.76e-14 &     7.48e-14 \\
16.0 &     2.75e-14 &     3.09e-14 &     3.53e-14 & &     3.23e-14 &     3.85e-14 &     4.72e-14 \\
\hline
\multicolumn{8}{c}{Cerro Tolonchar: 8995 used profiles, January 2006 -- February 2008} \\
\hline
 0.0 &     1.03e-13 &     1.24e-13 &     1.41e-13 & &     9.59e-14 &     1.17e-13 &     1.33e-13 \\
 0.5 &     8.53e-16 &     2.46e-15 &     4.17e-15 & &     1.15e-14 &     2.23e-14 &     4.61e-14 \\
 1.0 &     1.02e-15 &     1.33e-15 &     2.81e-15 & &     4.19e-15 &     7.76e-15 &     2.24e-14 \\
 2.0 &     3.17e-15 &     1.30e-14 &     4.32e-14 & &     1.23e-14 &     2.95e-14 &     6.76e-14 \\
 4.0 &     2.91e-14 &     3.96e-14 &     5.98e-14 & &     3.39e-14 &     4.68e-14 &     7.55e-14 \\
 8.0 &     2.14e-14 &     3.86e-14 &     7.46e-14 & &     3.65e-14 &     6.01e-14 &     8.67e-14 \\
16.0 &     3.07e-14 &     3.66e-14 &     4.61e-14 & &     3.38e-14 &     4.28e-14 &     6.01e-14 \\
\hline
\multicolumn{8}{c}{SPM: 13935 used profiles, October 2004 -- February 2008} \\
\hline
 0.0 &     2.06e-13 &     3.09e-13 &     5.55e-13 & &     1.94e-13 &     2.90e-13 &     5.07e-13 \\
 0.5 &     1.07e-15 &     2.35e-15 &     6.73e-15 & &     1.56e-14 &     2.75e-14 &     6.10e-14 \\
 1.0 &     2.52e-16 &     6.51e-16 &     1.33e-15 & &     4.04e-15 &     9.02e-15 &     2.56e-14 \\
 2.0 &     1.89e-16 &     3.80e-16 &     1.67e-15 & &     4.55e-15 &     1.03e-14 &     3.71e-14 \\
 4.0 &     9.77e-15 &     1.62e-14 &     2.80e-14 & &     1.71e-14 &     3.00e-14 &     5.49e-14 \\
 8.0 &     3.91e-15 &     1.38e-14 &     2.86e-14 & &     1.90e-14 &     3.75e-14 &     6.12e-14 \\
16.0 &     3.08e-14 &     3.62e-14 &     4.76e-14 & &    3.55e-14 &     4.47e-14 &     6.60e-14 \\
\hline
\multicolumn{8}{c}{13N Mauna Kea: 12180 used profiles, June 2005 -- February 2008} \\
\hline
 0.0 &     1.83e-13 &     2.85e-13 &     4.89e-13 & &     1.67e-13 &     2.56e-13 &     4.28e-13 \\
 0.5 &     1.45e-15 &     4.89e-15 &     1.22e-14 & &     1.78e-14 &     3.43e-14 &     6.49e-14 \\
 1.0 &     1.33e-15 &     2.36e-15 &     5.03e-15 & &     5.06e-15 &     1.02e-14 &     3.24e-14 \\
 2.0 &     2.93e-16 &     9.77e-16 &     3.62e-15 & &     4.75e-15 &     1.33e-14 &     3.21e-14 \\
 4.0 &     1.25e-14 &     1.95e-14 &     2.58e-14 & &     1.95e-14 &     3.03e-14 &     4.48e-14 \\
 8.0 &     4.30e-15 &     7.30e-15 &     1.29e-14 & &     1.67e-14 &     2.64e-14 &     4.46e-14 \\
16.0 &     1.89e-14 &     2.06e-14 &     2.34e-14 & &     2.40e-14 &     2.72e-14 &     3.48e-14 \\
\hline
\end{tabular}
}
\end{center}
\end{table}
\clearpage
\begin{table}
\begin{center}
{\scriptsize
\caption{ Mean and median $C_n^2~dh$ profiles constructed from 10\% of all profiles collected at each site,  
around the 25, 50 and 75 percentiles of the MASS $\theta_0$. The number of profiles used and time span covered by 
these are given for each site. }
\label{profileTHETA0seltable}
\begin{tabular}{c c c c c c c c c}
\hline
\hline
height & \multicolumn{3}{c}{median $C_n^2~dh ~[m^{1/3}]$}      & & \multicolumn{3}{c}{mean $C_n^2~dh ~[m^{1/3}] $} \\
$[$km$]$   &   25\%  $\theta_0$   &   50\% $\theta_0$               &  75\% $\theta_0$  & &   25\%  $\theta_0$   &  50\% $\theta_0$          &   75\% $\theta_0$ \\
\hline
\multicolumn{8}{c}{Cerro Tolar: 19657 used profiles, October 2003 -- April 2007} \\
\hline
 0.0 &     1.32e-13 &     1.20e-13 &     9.28e-14 & &     2.00e-13 &     1.82e-13 &     1.47e-13 \\
 0.5 &     7.43e-15 &     5.92e-15 &     9.65e-16 & &     4.45e-14 &     5.10e-14 &     4.33e-14 \\
 1.0 &     3.08e-15 &     3.86e-15 &     3.67e-15 & &     1.82e-14 &     2.52e-14 &     3.28e-14 \\
 2.0 &     2.17e-15 &     2.93e-15 &     8.35e-15 & &     2.39e-14 &     3.42e-14 &     5.18e-14 \\
 4.0 &     2.67e-14 &     3.42e-14 &     2.87e-14 & &     3.24e-14 &     4.39e-14 &     4.84e-14 \\
 8.0 &     7.49e-15 &     2.18e-14 &     6.76e-14 & &     1.53e-14 &     3.14e-14 &     7.11e-14 \\
16.0 &     2.68e-14 &     3.83e-14 &     5.47e-14 & &     2.52e-14 &     3.60e-14 &     5.28e-14 \\
\hline
\multicolumn{8}{c}{Cerro Armazones: 21238 used profiles, November 2004 -- February 2008} \\
\hline
 0.0 &     1.32e-13 &     1.23e-13 &     1.06e-13 & &     1.90e-13 &     1.76e-13 &     1.68e-13 \\
 0.5 &     1.53e-14 &     1.37e-14 &     5.12e-15 & &     6.45e-14 &     8.82e-14 &     7.45e-14 \\
 1.0 &     2.35e-15 &     2.60e-15 &     2.56e-15 & &     1.73e-14 &     2.25e-14 &     3.08e-14 \\
 2.0 &     4.86e-15 &     7.31e-15 &     1.17e-14 & &     2.05e-14 &     2.95e-14 &     3.97e-14 \\
 4.0 &     1.83e-14 &     2.26e-14 &     2.27e-14 & &     2.29e-14 &     3.00e-14 &     3.75e-14 \\
 8.0 &     1.24e-14 &     2.85e-14 &     7.16e-14 & &     1.75e-14 &     3.46e-14 &     7.57e-14 \\
16.0 &     2.25e-14 &     3.30e-14 &     4.86e-14 & &     2.15e-14 &     3.15e-14 &     4.69e-14 \\
\hline
\multicolumn{8}{c}{Cerro Tolonchar: 8995 used profiles, January 2006 -- February 2008} \\
\hline
 0.0 &     1.17e-13 &     1.12e-13 &     8.79e-14 & &     1.40e-13 &     1.32e-13 &     1.05e-13 \\
 0.5 &     7.59e-15 &     4.53e-15 &     1.17e-18 & &     4.20e-14 &     4.16e-14 &     3.11e-14 \\
 1.0 &     1.87e-15 &     2.42e-15 &     5.66e-16 & &     1.25e-14 &     2.15e-14 &     1.96e-14 \\
 2.0 &     7.00e-15 &     8.38e-15 &     2.16e-14 & &     3.18e-14 &     3.87e-14 &     6.41e-14 \\
 4.0 &     3.34e-14 &     4.58e-14 &     3.60e-14 & &     3.85e-14 &     5.44e-14 &     5.51e-14 \\
 8.0 &     1.09e-14 &     2.55e-14 &     9.83e-14 & &     1.79e-14 &     3.55e-14 &     1.02e-13 \\
16.0 &     2.56e-14 &     3.98e-14 &     5.88e-14 & &     2.42e-14 &     3.76e-14 &     5.66e-14 \\
\hline
\multicolumn{8}{c}{SPM: 13935 used profiles, October 2004 -- February 2008} \\
\hline
 0.0 &     2.39e-13 &     2.61e-13 &     3.22e-13 & &     3.44e-13 &     3.90e-13 &     5.36e-13 \\
 0.5 &     5.03e-15 &     4.08e-15 &     1.73e-15 & &     3.91e-14 &     5.42e-14 &     6.44e-14 \\
 1.0 &     7.48e-16 &     9.78e-16 &     5.68e-16 & &     8.99e-15 &     1.51e-14 &     3.12e-14 \\
 2.0 &     3.09e-16 &     3.93e-16 &     7.34e-16 & &     9.43e-15 &     1.85e-14 &     3.65e-14 \\
 4.0 &     1.27e-14 &     1.86e-14 &     2.07e-14 & &     1.95e-14 &     3.29e-14 &     4.58e-14 \\
 8.0 &     9.21e-16 &     1.01e-14 &     4.77e-14 & &     6.73e-15 &     2.00e-14 &     5.60e-14 \\
16.0 &     2.51e-14 &     3.89e-14 &     6.20e-14 & &     2.41e-14 &     3.72e-14 &     6.02e-14 \\
\hline
\multicolumn{8}{c}{13N Mauna Kea: 12180 used profiles, June 2005 -- February 2008} \\
\hline
 0.0 &     2.32e-13 &     2.44e-13 &     2.42e-13 & &     3.33e-13 &     3.69e-13 &     3.59e-13 \\
 0.5 &     7.29e-15 &     1.14e-14 &     4.59e-15 & &     4.10e-14 &     7.89e-14 &     1.12e-13 \\
 1.0 &     1.79e-15 &     2.68e-15 &     3.46e-15 & &     1.62e-14 &     2.76e-14 &     5.44e-14 \\
 2.0 &     5.80e-16 &     8.99e-16 &     1.10e-15 & &     1.02e-14 &     2.06e-14 &     3.38e-14 \\
 4.0 &     1.30e-14 &     2.12e-14 &     2.90e-14 & &     1.79e-14 &     2.95e-14 &     4.25e-14 \\
 8.0 &     1.48e-15 &     5.93e-15 &     2.39e-14 & &     4.89e-15 &     1.32e-14 &     3.51e-14 \\
16.0 &     1.50e-14 &     2.29e-14 &     3.70e-14 & &     1.43e-14 &     2.15e-14 &     3.47e-14 \\
\hline
\end{tabular}
}
\end{center}
\end{table}
\clearpage
\begin{table}
\begin{center}
{\scriptsize
\caption{ Mean and median $C_n^2~dh$ profiles constructed from 10\% of all profiles collected at each site,  
around the 25, 50 and 75 percentiles of the total $\tau_0$. The total $\tau_0$ was computed using $\tau_{0,\mathrm{MASS}}$
and adding the ground layer contribution. 
The number of profiles used and time span covered by these are given for each site. }
\label{profileTAU0seltable}
\begin{tabular}{c c c c c c c c c}
\hline
\hline
height & \multicolumn{3}{c}{median $C_n^2~dh ~[m^{1/3}]$}      & & \multicolumn{3}{c}{mean $C_n^2~dh ~[m^{1/3}] $} \\
$[$km$]$   &   25\%  $\tau_0$   &   50\% $\tau_0$               &   75\% $\tau_0$  & &   25\%  $\tau_0$   &  50\% $\tau_0$          &   75\% $\tau_0$ \\
\hline
\multicolumn{8}{c}{Cerro Tolar: 19402 used profiles, October 2003 -- April 2007} \\
\hline
 0.0 &     1.41e-13 &     1.11e-13 &     7.97e-14 & &     2.02e-13 &     1.76e-13 &     1.29e-13 \\
 0.5 &     9.42e-15 &     4.13e-15 &     7.54e-16 & &     5.24e-14 &     4.72e-14 &     3.42e-14 \\
 1.0 &     4.12e-15 &     3.02e-15 &     3.07e-15 & &     3.26e-14 &     2.32e-14 &     2.42e-14 \\
 2.0 &     5.87e-15 &     2.69e-15 &     4.23e-15 & &     3.86e-14 &     3.50e-14 &     4.38e-14 \\
 4.0 &     2.94e-14 &     2.81e-14 &     2.86e-14 & &     3.87e-14 &     4.00e-14 &     4.88e-14 \\
 8.0 &     8.51e-15 &     2.39e-14 &     6.83e-14 & &     1.98e-14 &     3.63e-14 &     7.49e-14 \\
16.0 &     2.80e-14 &     3.64e-14 &     4.83e-14 & &     3.17e-14 &     4.01e-14 &     5.46e-14 \\
\hline
\multicolumn{8}{c}{Cerro Armazones: 20324 used profiles, November 2004 -- February 2008} \\
\hline
 0.0 &     1.31e-13 &     1.12e-13 &     1.01e-13 & &     1.87e-13 &     1.58e-13 &     1.58e-13 \\
 0.5 &     1.36e-14 &     1.25e-14 &     6.33e-15 &  &    6.86e-14 &     7.97e-14 &     7.56e-14 \\
 1.0 &     1.99e-15 &     2.35e-15 &     2.35e-15 & &     1.76e-14 &     2.35e-14 &     2.69e-14 \\
 2.0 &     4.75e-15 &     7.67e-15 &     1.12e-14 & &     1.98e-14 &     2.89e-14 &     3.93e-14 \\
 4.0 &     1.59e-14 &     2.12e-14 &     2.49e-14 & &     2.21e-14 &     2.91e-14 &     3.64e-14 \\
 8.0 &     1.13e-14 &     2.84e-14 &     6.51e-14 & &     1.83e-14 &     3.73e-14 &     7.31e-14 \\
16.0 &     2.36e-14 &     3.06e-14 &     3.95e-14 & &     2.67e-14 &     3.50e-14 &     4.60e-14 \\
\hline
\multicolumn{8}{c}{Cerro Tolonchar: 8546 used profiles, January 2006 -- February 2008} \\
\hline
 0.0 &     1.20e-13 &     1.01e-13 &     6.26e-14 & &     1.43e-13 &     1.23e-13 &     7.65e-14 \\
 0.5 &     4.47e-15 &     1.49e-15 &     7.96e-18 & &     4.05e-14 &     3.79e-14 &     3.52e-14 \\
 1.0 &     1.53e-15 &     1.60e-15 &     8.31e-16 & &     1.57e-14 &     1.91e-14 &     2.66e-14 \\
 2.0 &     6.94e-15 &     8.55e-15 &     2.46e-14 & &     3.28e-14 &     4.06e-14 &     6.38e-14 \\
 4.0 &     2.98e-14 &     3.63e-14 &     5.92e-14 & &     3.70e-14 &     4.54e-14 &     7.11e-14 \\
 8.0 &     1.57e-14 &     3.91e-14 &     1.10e-13 &  &    2.82e-14 &     5.07e-14 &     1.12e-13 \\
16.0 &     3.14e-14 &     3.71e-14 &     5.35e-14 &  &    3.55e-14 &     4.29e-14 &     6.85e-14 \\
\hline
\multicolumn{8}{c}{SPM: 13749 used profiles, October 2004 -- February 2008} \\
\hline
 0.0 &     2.29e-13 &     2.95e-13 &     3.87e-13 & &     3.24e-13 &     4.26e-13 &     6.38e-13 \\
 0.5 &     5.56e-15 &     3.70e-15 &     1.40e-15 & &     4.61e-14 &     5.30e-14 &     7.65e-14 \\
 1.0 &     8.20e-16 &     8.35e-16 &     2.40e-16 & &     1.30e-14 &     1.82e-14 &     3.75e-14 \\
 2.0 &     3.71e-16 &     3.99e-16 &     4.32e-16 & &     1.47e-14 &     2.52e-14 &     3.23e-14 \\
 4.0 &     1.15e-14 &     1.67e-14 &     1.91e-14 & &     2.46e-14 &     3.46e-14 &     4.22e-14 \\
 8.0 &     1.32e-15 &     8.79e-15 &     5.51e-14 & &     1.14e-14 &     2.45e-14 &     6.61e-14 \\
16.0 &     2.77e-14 &     3.63e-14 &     5.53e-14 & &     3.23e-14 &     4.23e-14 &     6.58e-14 \\
\hline
\multicolumn{8}{c}{13N Mauna Kea: 12030 used profiles, June 2005 -- February 2008} \\
\hline
 0.0 &     2.46e-13 &     2.50e-13 &     2.21e-13 & &     3.70e-13 &     3.95e-13 &     3.74e-13 \\
 0.5 &     1.04e-14 &     1.06e-14 &     2.51e-15 & &     5.74e-14 &     8.16e-14 &     1.55e-13 \\
 1.0 &     2.50e-15 &     2.88e-15 &     3.99e-15 & &     2.68e-14 &     4.07e-14 &     7.53e-14 \\
 2.0 &     8.59e-16 &     8.83e-16 &     2.31e-15 & &     1.97e-14 &     2.31e-14 &     4.04e-14 \\
 4.0 &     1.52e-14 &     2.20e-14 &     3.55e-14 & &     2.30e-14 &     3.18e-14 &     5.17e-14 \\
 8.0 &     2.84e-15 &     9.31e-15 &     4.18e-14 & &     9.47e-15 &     2.17e-14 &     5.48e-14 \\
16.0 &     1.75e-14 &     2.35e-14 &     3.68e-14 & &     2.03e-14 &     2.77e-14 &     4.22e-14 \\
\hline
\end{tabular}
}
\end{center}
\end{table}

\end{document}